\documentclass[11pt,a4paper]{article}
\pdfoutput=1
\usepackage{jstylemarg}
\usepackage[utf8]{inputenc}
\usepackage{amsmath}
\usepackage{mathtools}
\usepackage{tikz}
\usepackage{empheq}
\usepackage{enumitem}
\usepackage{graphics}
\usepackage{wrapfig}
\usepackage{caption}
\usepackage{xcolor}
\usepackage{framed}
\definecolor{shadecolor}{gray}{0.925}

\newcommand{\bi}{\begin{itemize}}
\newcommand{\ei}{\end{itemize}}
\newcommand{\bea}{\begin{align}}
\newcommand{\eea}{\end{align}}
\newcommand{\be}{\begin{equation}}
\newcommand{\ee}{\end{equation}}

\newcommand{\Tr}{{\text{Tr}}}

\newcommand{\pl}{{\partial}}

\newcommand{\tcb}{\textcolor{blue}}

\makeatletter
\renewcommand*\env@matrix[1][\arraystretch]{%
  \edef\arraystretch{#1}%
  \hskip -\arraycolsep
  \let\@ifnextchar\new@ifnextchar
  \array{*\c@MaxMatrixCols c}}
\makeatother

\author[\ensuremath{\ell}]{Charlotte SLEIGHT\footnote{Also at the Universit\'e Libre de Bruxelles and International Solvay Institutes, Belgium.}}
\author[\ensuremath{\mathsf{s}},\ensuremath{\mathsf{t}},\ensuremath{\mathsf{u}}]{\quad Massimo TARONNA}

\affiliation[\ensuremath{\ell}]{School of Natural Sciences, Institute for Advanced Study,\\
1 Einstein Drive, Princeton, NJ 08540}

\affiliation[\ensuremath{\mathsf{s}}]{Department of Physics, Princeton University,\\
Jadwin Hall, Princeton, NJ 08544}

\affiliation[\ensuremath{\mathsf{t}}]{Dipartimento di Fisica ``Ettore Pancini'', Universit\`a degli Studi di Napoli Federico II, \\Monte S. Angelo, Via Cintia, 80126 Napoli, Italy}

\affiliation[\ensuremath{\mathsf{u}}]{INFN, Sezione di Napoli, Monte S. Angelo, Via Cintia, 80126 Napoli, Italy}

\emailAdd{csleight@ias.edu, massimo.taronna@unina.it}

\title{\centering\huge{The Unique Polyakov Blocks}}

\abstract{In this work we present a closed form expression for Polyakov blocks in Mellin space for arbitrary spin and scaling dimensions. We provide a prescription to fix the contact term ambiguity uniquely by reducing the problem to that of fixing the contact term ambiguity at the level of cyclic exchange amplitudes --- defining cyclic Polyakov blocks --- in terms of which any fully crossing symmetric correlator can be decomposed. We also give another, equivalent, prescription which does not rely on a decomposition into cyclic amplitudes and we underline the relation between cyclic amplitudes and dispersion relations in Mellin space. We extract the OPE data of double-twist operators in the direct channel expansion of the cyclic Polyakov blocks using and extending the analysis of \cite{Sleight:2018epi,Sleight:2018ryu} to include contributions that are non-analytic in spin. The relation between cyclic Polyakov blocks and analytic Bootstrap functionals is underlined.}

\begin{document}

\maketitle

\section{The problem and its resolution}
\label{sec::problemandresolution}

The simplest solutions to the crossing equation in Conformal Field Theory (CFT) that are single-valued in in the Euclidean region are given by tree-level exchange amplitudes in the dual anti-de Sitter (AdS) space, which have come to be known as Polyakov blocks \cite{Gopakumar:2016wkt}. It is well known that it is possible to bootstrap the Witten diagram for the exchange of a field with spin-$\ell$ and mass $m^2R^2=\Delta\left(\Delta-d\right)-\ell$ simply by requiring \cite{ElShowk:2011ag,Costa:2012cb,Costa:2014kfa,Alday:2017gde}: Crossing symmetry, the presence of a conformal block with scaling dimension $\Delta$ and spin $\ell$ in the OPE decomposition and Euclidean single-valuedness. This bootstrap problem is however subject to an ambiguity that is parameterised by solutions to crossing with finite support in spin. These are in one-to-one correspondence with bulk contact terms which, for spinning internal legs, have a better Regge behaviour than the full exchange solution itself. Naively one might expect that there does not exist a canonical choice of contact terms for Polyakov blocks. The presence of this ambiguity however poses a technical problem for the implementation of the Polyakov bootstrap \cite{Polyakov:1974gs,Gopakumar:2018xqi}, which is related to the fact that a given choice for the contact terms might not be compatible with convergence of the corresponding expansion in Polyakov blocks. Although this problem is expected to be technical in nature, since solutions to crossing are not subject to such ambiguities, it is imperative that it is resolved. So far a complete satisfactory solution valid in general dimensions has not been found.\footnote{See however \cite{Mazac:2016qev,Mazac:2018qmi,Mazac:2018ycv,Paulos:2019gtx} for some key ideas in this direction.}

In this work we propose a prescription for the contact term ambiguity, which is motivated by flat space scattering amplitudes and string theory. As we shall demonstrate in the following, this prescription allows to uniquely fix the entire contact term ambiguity in a simple manner.

Let us consider the analogous problem in flat space at tree-level. In string theory, a related problem would be to obtain the following decomposition of the Shapiro-Virasoro amplitude:
\begin{equation}\label{SVamplitude}
\frac{\Gamma(-1-\tfrac{\alpha^\prime}4 s)\Gamma(-1-\tfrac{\alpha^\prime}4 t)\Gamma(-1-\tfrac{\alpha^\prime}4 u)}{\Gamma(2+\tfrac{\alpha^\prime}4 s)\Gamma(2+\tfrac{\alpha^\prime}4 t)\Gamma(2+\tfrac{\alpha^\prime}4 u)}=\sum_{n,\ell}a_{n,\ell}P_{n,{\ell}}(s,t,u)\,,
\end{equation}
where $P_{n,\ell}(s,t,u)$ is the crossing symmetric sum of exchanges for a string excitation of spin ${\ell}$ and mass parameterised by the Regge trajectory $n$, which is symmetric in the Mandelstam invariants $s$, $t$ and $u$.\footnote{In principle there are contributions from string excitations of mixed symmetry which render the above problem technically complicated. For the following discussion such technical complications will not play a role.} In particular:
\begin{align}\label{Polyakov_flat_SV}
    P_{n,\ell}(s,t,u)\sim\frac{\# (s-u)^{\ell}+\ldots}{\alpha^\prime t+4(1-n)}+(\sf{s}-\text{channel})+(\sf{u}-\text{channel}).
\end{align}
We see that, contrary to the full Shapiro-Virasoro amplitude, a single crossing symmetric building block is not well behaved in the Regge limit $s \to \infty$ and $t$ fixed. We are therefore still free to add contact terms to $P_{n,\ell}(s,t,u)$ that are polynomials of degree $\ell-1$ in $s$, $t$ and $u$, which do not modify its leading Regge behaviour. The above decomposition problem is therefore ambiguous at best. While at the level of the full Shapiro-Virasoro amplitude a resolution to this problem appears unclear, it is instructive to consider open String Theory where the natural crossing symmetric object is given by the Veneziano amplitude. The key observation is that a given cyclicly-ordered Veneziano amplitude, say,
\begin{equation}
  \frac{\Gamma(-1-\tfrac{\alpha^\prime}2 t)\Gamma(-1-\tfrac{\alpha^\prime}2 u)}{\Gamma(-2+\tfrac{\alpha^\prime}2 s)},
\end{equation}
is crossing symmetric only with respect to two channels, in this case ${\sf t}$ and ${\sf u}$, so that we may consider a decomposition into cyclic exchange building blocks $\bar{P}_{n,\ell}(s|t,u)$ which are crossing symmetric with respect to those channels.
Contrary to the fully crossing symmetric building blocks \eqref{Polyakov_flat_SV}, the cyclic building blocks can be fixed by the requirement that they are suppressed in the Regge limits $t\to \infty$ or $u\to\infty$ at fixed $s$, meaning that the residues of the poles in  $t$ and $u$ are functions of $s$ only:
\begin{equation}\label{Polyakov_FS_V}
    \bar{P}_{n,\ell}(s|t,u)=\frac{\# s^\ell+\ldots}{\alpha^\prime t+2(1-n)}+\frac{\# s^\ell+\ldots}{\alpha^\prime u+2(1-n)}\,.
\end{equation}
With this prescription, we are setting to zero any additional contact terms polynomial in $t$ and $u$, which would not be suppressed in the above pair of Regge limits. Furthermore, under these boundary conditions, the residues are uniquely fixed by factorisation in terms of flat space partial waves.
This gives an unambiguous definition of the cyclicly-ordered crossing symmetric blocks \eqref{Polyakov_FS_V}. An amplitude $\mathcal{A}(s,t,u)$ that is crossing symmetric with respect to the ${\sf s}$-, ${\sf t}$- and ${\sf u}$-channels, such as the Shapiro-Virasoro amplitude \eqref{SVamplitude}, then admits a well defined decomposition in terms of a single set of cyclic Polyakov blocks
\begin{align}\label{flatcycdecomp}
    \mathcal{A}(s,t,u)=\sum a_{n,\ell}\bar{P}_{n,\ell}(s|t,u)\,,
\end{align}
so that manifest crossing symmetry with respect to all channels is recovered only after summing over the spectrum.
 
The above solution for the contact term ambiguity in flat space can be naturally extended to AdS$_{d+1}$ by working in Mellin space. It has been argued \cite{Mack:2009mi,Penedones:2010ue,Fitzpatrick:2011ia} that the AdS analogue of the flat space scattering amplitude is the so-called \emph{Mellin amplitude} $\mathcal{M}(s,t)$, which is defined as the Mellin transform of the CFT correlation function,
 \begin{align}\label{Mellinrepresentation}
    \mathcal{A}(u,v)=\int_{-i\infty}^{+i\infty}\frac{ds dt}{(4\pi i)^2}\,u^{t/2}v^{-(s+t)/2}\,\rho(s,t)\mathcal{M}(s,t)\,,
\end{align}
where $u$ and $v$ are the usual cross ratios and, for external operators with equal scaling dimension $\Delta$, we have\footnote{For generic external legs of twist $\tau_i$, this reads: 
\begin{align}\label{rhogen}
    \rho(s,t)=\Gamma \left(\tfrac{s+t}{2}\right) \Gamma \left(\tfrac{-s-\tau_1+\tau_2}{2}\right) \Gamma \left(\tfrac{-s+\tau_3-\tau_4}{2}\right) \Gamma \left(\tfrac{-t+\tau_1+\tau_2}{2}\right) \Gamma \left(\tfrac{-t+\tau_3+\tau_4}{2}\right) \Gamma \left(\tfrac{s+t+\tau_1-\tau_2-\tau_3+\tau_4}{2}\right)\,.
\end{align}
}
 \begin{align}\label{rhouequal}
    \rho_{\tau_i}(s,t)=\Gamma \left(-\frac{s}{2}\right)^2 \Gamma \left(\frac{s+t}{2}\right)^2 \Gamma \left(\frac{2 \Delta -t}{2}\right)^2.
\end{align}
The Mellin variables $s$ and $t$ are analogues of the Mandelstam invariants in flat space, where $s+t+u=2\Delta$.\footnote{In this work we follow the conventions of \cite{Costa:2012cb}. However, we choose to refer to the $\sf{s}$, $\sf{t}$ and $\sf{u}$-channels in a canonical way, so that in the conventions of \cite{Costa:2012cb} poles in the variables $t$ correspond to $\sf{s}$-channel exchanges. This is made clear in equation \eqref{channel_Mellin}.} The exchange of a spin-$\ell$ field, say, in the ${\sf s}$-channel takes the following simple form \cite{Costa:2012cb} 
 \begin{align}\label{SchExch}
    \mathcal{E}^{(\sf{s})}_{\tau,\ell}(u,v)=\int^{+i\infty}_{-i\infty}\frac{ds dt}{(4\pi i)^2}\,u^{t/2}v^{-(s+t)/2}\,\rho(s,t)\underbrace{\left[\left(\sum_{m=0}^\infty\frac{\mathcal{Q}_{\tau,\ell|m}(s)}{t-\tau-2m}\right)+p_{\ell-1}(s,t)\right]}_{\mathcal{E}_{\tau,\ell}^{(\bf{s})}(s,t)}\,.
\end{align}
Similar to flat space exchanges,  the exchanged single-particle state corresponds to single poles in the Mellin variable $t$, whose residues are given by the kinematic polynomials of degree $\ell$ in the Mellin variable $s$, ${\mathcal{Q}_{\tau,\ell|m}(s)=s^\ell+O\left(s^{\ell-1}\right)}$. These are reviewed in Appendix \ref{Hahn} and are fixed by factorisation. The function $p_{\ell-1}(s,t)$ is a polynomial of degree $\ell-1$ in the Mellin variables $s$ and $t$ and parameterises the contact term ambiguity. In particular, such a polynomial does not modify the leading behaviour of the exchange \eqref{SchExch} in the Regge limit $s \to \infty$ and $t$ fixed\footnote{This definition of the Regge limit for Mellin amplitudes was given in \cite{Costa:2012cb} where, going to momentum space and defining the corresponding Mandelstam invariants, the above Regge limit of the Mellin amplitude is equivalent to the usual Regge limit defined in terms of the Mandelstam invariants.} and corresponds to the freedom to add on-shell vanishing terms (also known as improvements) to the cubic couplings.
 
Given the similarities between Mellin amplitudes and flat space amplitudes, we can immediately extend the prescription for the contact ambiguity in flat space to AdS. Accordingly, one considers cyclicly ordered exchange amplitudes $\mathcal{E}_{\tau,\ell}^{(1\sigma(2)\sigma(3)\sigma(4))}$ in AdS where, working with unequal external legs, the full crossing symmetric exchange amplitude reads
\begin{align}\label{co_sum}
    \mathcal{E}_{\tau,\ell}&=\sum_\sigma \Tr\left[T^{a_1}T^{a_{\sigma(2)}}T^{a_{\sigma(3)}}T^{a_{\sigma(4)}}\right]\mathcal{E}_{\tau,\ell}^{(1\sigma(2)\sigma(3)\sigma(4))},
\end{align}
with the usual trace factor. The case of identical external legs is associated to the singlet sector
of the colour group. For odd spins $\ell$, the ${\sf s}$-channel exchange reads:
\begin{equation}
\widehat{\mathcal{E}}_{\tau,\ell}^{(\sf{s})}=\Tr\Big([T^{a_1},T^{a_{2}}][T^{a_{3}},T^{a_{4}}]\Big)\mathcal{E}_{\tau,\ell}^{(\sf{s})},
\end{equation}
while for even spins we have:
\begin{equation}\label{EvenS}
\widehat{\mathcal{E}}_{\tau,\ell}^{(\sf{s})}=\Tr\Big(\{T^{a_1},T^{a_{2}}\}\{T^{a_{3}},T^{a_{4}}\}\Big)\mathcal{E}_{\tau,\ell}^{(\sf{s})}\,,
\end{equation}
so that the cyclic exchange amplitudes are given in terms of the $\sf{s}$-, $\sf{t}$- and $\sf{u}$-channel exchanges as:
\begin{subequations}
\begin{align}
    \mathcal{E}_{\tau,\ell}^{(1234)}&=\mathcal{E}_{\tau,\ell}^{(\sf{s})}+\mathcal{E}_{\tau,\ell}^{(\sf{u})}\,,\\
    \mathcal{E}_{\tau,\ell}^{(1342)}&=(-1)^\ell\left[\mathcal{E}_{\tau,\ell}^{(\sf{s})}+\mathcal{E}_{\tau,\ell}^{(\sf{t})}\right]\,,\\
    \mathcal{E}_{\tau,\ell}^{(1423)}&=\mathcal{E}_{\tau,\ell}^{(\sf{t})}+(-1)^\ell\mathcal{E}_{\tau,\ell}^{(\sf{u})}\,.
\end{align}
\end{subequations}
These are the AdS analogues of the flat space cyclic building blocks \eqref{Polyakov_FS_V}. To make the connection with the flat space analysis more transparent, it is convenient to define the following canonical Mellin variables:
\begin{align}\label{channel_Mellin}
    S&=t\,,& T&=s+2\Delta\,,& U&=-s-t+2\Delta\,,
\end{align}
where $S+T+U=4\Delta$, which have the property of being mapped into each other under crossing. In terms of these, the cyclic exchange amplitudes can be expressed in the following compact form:
\begin{subequations}\label{cyclic_exch}
\begin{align}
    \mathcal{E}_{\tau,\ell}^{(1234)}&=\sum_{m=0}^\infty \left[\frac{\mathcal{Q}_{\tau,\ell|m}(T-2\Delta)}{S-\tau-2m}+\frac{\mathcal{Q}_{\tau,\ell|m}(T-2\Delta)}{U-\tau-2m}\right]+p_{\ell-1}(S,U)\,,\label{exch1}\\
    \mathcal{E}_{\tau,\ell}^{(1423)}&=\sum_{m=0}^\infty \left[\frac{\mathcal{Q}_{\tau,\ell|m}(S-2\Delta)}{U-\tau-2m}+\frac{\mathcal{Q}_{\tau,\ell|m}(S-2\Delta)}{T-\tau-2m}\right]+p_{\ell-1}(U,T),\,\\
    \mathcal{E}_{\tau,\ell}^{(1342)}&=\sum_{m=0}^\infty \left[\frac{\mathcal{Q}_{\tau,\ell|m}(U-2\Delta)}{T-\tau-2m}+\frac{\mathcal{Q}_{\tau,\ell|m}(U-2\Delta)}{S-\tau-2m}\right]+p_{\ell-1}(T,S)\,,
\end{align}
\end{subequations}
where the residues of the single poles are a function of a single Mellin variable. The contact term ambiguity is parametrised by a symmetric polynomial $p_{\ell-1}(X,Y)=p_{\ell-1}(Y,X)$ of degree $\ell-1$ in $X$ and $Y$ together (see Appendix \ref{ContactAmb}). 

The contact term ambiguity can thus be fixed uniquely by requiring that each cyclic exchange amplitude is suppressed in the corresponding pair of Regge limits,\footnote{E.g. for \eqref{exch1} these would be $S \to \infty$ or $U \to \infty$ with $T$ fixed.} just like in the flat space analysis. This condition is satisfied by the single pole terms, but clearly violated by a non-trivial $p_{\ell-1}(X,Y)$ which must therefore be set to zero. With this prescription, the cyclicly ordered Polyakov blocks are given unambiguously by:\footnote{Although in this discussion we considered the case of external operators with equal scaling dimension $\Delta$, it trivially to extends to operators of unequal scaling dimensions with measure \eqref{rhogen}.
}
{\allowdisplaybreaks
\begin{shaded}
\begin{subequations}\label{cyclic_exch_fin} \noindent \emph{Unique Cyclicly Ordered Polyakov Blocks.}
\begin{align}
    \mathcal{E}_{\tau,\ell}^{(1234)}&=\sum_{m=0}^\infty \left[\frac{\mathcal{Q}_{\tau,\ell|m}(T-2\Delta)}{S-\tau-2m}+\frac{\mathcal{Q}_{\tau,\ell|m}(T-2\Delta)}{U-\tau-2m}\right]\,,\label{exch1b}\\
    \mathcal{E}_{\tau,\ell}^{(1423)}&=\sum_{m=0}^\infty \left[\frac{\mathcal{Q}_{\tau,\ell|m}(S-2\Delta)}{U-\tau-2m}+\frac{\mathcal{Q}_{\tau,\ell|m}(S-2\Delta)}{T-\tau-2m}\right],\,\\
    \mathcal{E}_{\tau,\ell}^{(1342)}&=\sum_{m=0}^\infty \left[\frac{\mathcal{Q}_{\tau,\ell|m}(U-2\Delta)}{T-\tau-2m}+\frac{\mathcal{Q}_{\tau,\ell|m}(U-2\Delta)}{S-\tau-2m}\right]\,.
\end{align}
\end{subequations}
\end{shaded}}
\noindent Since each cyclic Polyakov block is well-behaved in the corresponding Regge limits, the decomposition of crossing symmetric amplitudes in terms of these building blocks is well defined. 

It is instructive to note that the above form of cyclic Polyakov blocks can also be derived from a dispersion relation argument for the Mellin amplitude assuming that the amplitude is suppressed in the Regge limit at $T$ fixed:
\begin{align}
\mathcal{M}(S,T,U)=\oint_S dz\,\frac1{z-S}\,\mathcal{M}(z,T,-T-z+4\Delta)=\sum_{\tau,\ell}a_{\tau,\ell}\mathcal{E}_{\tau,\ell}^{(1234)}\,.
\end{align}
To get the above relations we simply closed the integration contour on the singularities of $\mathcal{M}(z,T,-T-z+4\Delta)$ which reproduce precisely the $\mathsf{s}$ and $\mathsf{u}$ singularities present in our Polyakov block $\mathcal{E}_{\tau,\ell}^{(1234)}$ defined in \eqref{cyclic_exch_fin}.
This works in the same way also in flat space. The cyclic amplitude is naturally generated. A better dispersion relation which takes into account the behaviour of the Mellin amplitude at infinity \cite{Costa:2012cb} can be obtained considering subtractions which improve the behaviour of the Mellin amplitude at infinity. A clever subtraction can be performed recalling that the Mellin amplitude has to have a zero at the location of the leading Regge trajectory \cite{Gopakumar:2016cpb,Penedones:2019tng}. In this case indeed it is natural to write down the following subtracted dispersion relation:
\begin{align}
    \mathcal{M}(S,T,U)=(S-2\Delta)(U-2\Delta)\sum_{\tau,\ell}a_{\tau,\ell}\sum_{m=0}^\infty \frac{1}{(\tau-2\Delta+2m) (T-2 \Delta+\tau  +2 m)}\mathcal{E}_{\tau,\ell,|m}^{(1234)}\,,
\end{align}
where we used that $\mathcal{M}(2\Delta,T,2\Delta-T)=\mathcal{M}(2\Delta-T,T,2\Delta)=0$.
Similar subtracted dispersion relations can be obtained including in the Mellin amplitude more poles from the overall $\Gamma$-function factor leading to:
\begin{align}
    \frac{\mathcal{M}(S,T,U)}{(2 \Delta -S)_\alpha (2 \Delta -T)_\alpha (2 \Delta -U)_\alpha}=\sum_{\tau,\ell}a_{\tau,\ell}\sum_{m=0}^\infty\frac{\mathcal{E}_{\tau,\ell|m}^{(1234)}(S,U)}{(2 \Delta -T)_{\alpha } (-2 n+2 \Delta -\tau )_{\alpha } (2 n+T-2 \Delta +\tau )_{\alpha }}\,\,.
\end{align}
The above subtracted dispersion relation make manifest the Polyakov bootstrap conditions satisfied by the Mellin amplitude but because of the nature of the subtraction crossing is not manifest among all possible channels (for instance between $S$ and $T$ channels). This naturally leads to sum rules (see e.g. \cite{Penedones:2019tng} for related results).

To conclude, let us also comment on the fact that in the case of identical external legs the above solution of the contact term ambiguity for the cyclic amplitude also fixes the contact term ambiguity for a putative fully crossing symmetric Polyakov block, which is trivially captured by the singlet sector of the colour group \eqref{co_sum}. In this case the unique fully crossing symmetric Polyakov block is simply given, with a certain normalisation for the color factors, by
\begin{equation}\label{PBidleg}
     \mathcal{E}_{\tau,\ell}=\frac{1}{2}\left[ \mathcal{E}_{\tau,\ell}^{(1234)}+ \mathcal{E}_{\tau,\ell}^{(1423)}+\mathcal{E}_{\tau,\ell}^{(1342)}\right].
\end{equation}
In section \ref{sec::idelgs} we show how the contact term ambiguity in this case can be equivalently fixed without having to introduce cyclicly ordered exchange amplitudes, by identifying universal terms in exchange diagrams which are polynomial in the Mellin variables but cannot be removed by adding contact terms. That the corresponding contact term ambiguity is fixed in the same way as it is in \eqref{PBidleg} via the cyclic Polyakov blocks \eqref{cyclic_exch_fin} is suggestive that CFT correlators could still admit a decomposition in terms of fully crossing symmetric Polyakov blocks \eqref{PBidleg}, though this question is not the subject of the present work and convergence of the corresponding expansion might only be conditional.

Having provided an unambiguous definition of (cyclic) Polyakov blocks, in section \ref{sec::CBdecomp} we derive their conformal block decomposition and the corresponding OPE data of double-twist operators. In the process we clarify some computational subtleties in extracting the contributions to the OPE data of low spin double-twist operators. We furthermore show that the cyclic Polyakov blocks \eqref{cyclic_exch_fin} can be regarded as generating functions for analytic Bootstrap functionals, so that our prescription to uniquely fix cyclic Polyakov blocks translates into a prescription to define analytic Bootstrap functionals in general $d$ for generic spin.

Various technical details and definitions are relegated to the appendices.

\paragraph{Note added:} In the final stage of preparing this draft we became aware of the work \cite{Mazac:2019shk}, presenting results in partial overlap with ours. However, our general explicit result \eqref{cyclic_exch_fin} for the Polyakov blocks and the approach to obtain them are new. In \cite{Mazac:2019shk} explicit examples of unique Polyakov blocks were constructed in the $\ell=0$ and $\ell=1$ cases. M.T. thanks X.Z. for discussions in spring 2019, in which some aspects of our results were communicated. 

The recent interesting work \cite{Ferrero:2019luz} fixes the contact term ambiguity in CFT$_1$.

\section{An alternative derivation for identical external legs}
\label{sec::idelgs}

Starting from the Mellin representation \eqref{SchExch} of an $\sf{s}$-channel exchange, one can easily generate the corresponding crossing symmetric solution by adding up the different channels. To this end, it is useful to introduce operations that interchange the external legs of a given correlator, which at the level of the Mellin amplitude correspond to the following transformations:
\begin{subequations}\label{perm}
\begin{align}
f(s,t)\Big|_{1\to2,2\to1}=f(s,t)\Big|_{3\to4,4\to3}&=f(-s-t,t)\,,\\
f(s,t)\Big|_{2\to4,4\to2}=f(s,t)\Big|_{1\to3,3\to1}&=f(s,-s-t+2\Delta)\,,\\
f(s,t)\Big|_{1\to4,4\to1}=f(s,t)\Big|_{2\to3,3\to2}&=f(t-2\Delta,s+2\Delta)\,.
\end{align}
\end{subequations}
With the above operations one can define various projectors. The simplest are the projectors onto the symmetric and anti-symmetric part of the ${\sf s}$-channel amplitude under the exchange of the legs $1$ and $2$ or $3$ and $4$
\begin{subequations}
\begin{align}
    \mathcal{S}[f(s,t)]&=\frac{f(s,t)+f(-s-t,t)}2\,,\\
    \mathcal{A}[f(s,t)]&=\frac{f(s,t)-f(-s-t,t)}2\,.
\end{align}
\end{subequations}
These distinguish even and odd spins since odd-exchanges are anti-symmetric and even-exchange are symmetric. We can also define an operation which takes a function $f(s,t)$ and generates from it a crossing-symmetric object:
\begin{equation}
    \mathfrak{C}[f(s,t)]=f(s,t)+f(t-2\Delta,s+2\Delta)+f(s,-s-t+2\Delta)\,.
\end{equation}
The above operation can then be used to define additional crossing symmetric projectors as composition of $\mathfrak{C}$, $\mathcal{S}$ and $\mathcal{A}$. 

For identical external legs the full crossing symmetric exchange amplitude $\mathcal{E}_{\tau,\ell}(s,t)$ is completely symmetric under any permutation of the external legs. It is therefore generated from the ${\sf s}$-channel exchange \eqref{SchExch} by symmetrising the two pairs of external legs 12 and 34 with $\mathcal{S}$, then applying $\mathfrak{C}$:
\begin{equation}\label{idCSexc}
     \mathcal{E}_{\tau,\ell}(s,t)=\mathfrak{C}\circ\mathcal{S}\left[\mathcal{E}^{(\sf{s})}_{\tau,\ell}(s,t)\right].
\end{equation}
This operator moreover organises polynomials in $s$ and $t$ that can appear in the exchange amplitude into two types:
\begin{enumerate}
    \item \underline{Polynomials in the image of $\mathfrak{C}\circ\mathcal{S}$}, where each Eigenvector corresponds to a solution of the crossing equation that has finite support in spin i.e. a contact amplitude. The contact term ambiguity is precisely the freedom to add such Eigenvectors up to degree $\ell-1$ in $s$ and $t$ to the exchange amplitude.
    \item  \underline{Polynomials that belong to the kernel $\ker\mathfrak{C}\circ\mathcal{S}$} instead do not correspond to contact amplitudes. As we shall see explicitly, they are a universal feature of exchange solutions to crossing, which means that such polynomials cannot be removed by the freedom to add contact terms.
\end{enumerate}
\noindent The operation $\mathfrak{C}\circ\mathcal{S}$ therefore conveniently disentangles the contact term ambiguity from exchange solutions to crossing. It therefore provides a minimal way to fix the contact term ambiguity, by requiring that all polynomial terms in the exchange amplitude \eqref{idCSexc} belong to $\ker\mathfrak{C}\circ\mathcal{S}$. To explore how this condition is implemented, let us for the moment set the polynomial $p_{\ell-1}\left(s,t\right)$ in the ${\sf s}$-channel exchange \eqref{SchExch} to zero --- which we are free to do whilst the contact term ambiguity is unfixed. The corresponding crossing symmetric exchange amplitude is:
\begin{multline}\label{crossExch}
    \mathcal{E}_{\tau,\ell}(s,t)=\mathfrak{C}\circ\mathcal{S}\left[\mathcal{E}^{(\sf{s})}_{\tau,\ell}(s,t)\right]\\
    =\sum_{m=0}^\infty\left(\tfrac{(2 \Delta-2 \tau -4 m-s )\mathcal{Q}_{\tau,\ell|m}(s)}{2 (t-\tau-2m ) (u-\tau-2 m )}+\tfrac{(4 \Delta-2 \tau -4 m+u )\mathcal{Q}_{\tau,\ell|m}(u-2\Delta)}{2 (t-\tau-2m ) (s+2 \Delta-\tau -2 m )}+\tfrac{(4 \Delta-2 \tau -4 m-t )\mathcal{Q}_{\tau,\ell|m}(t-2 \Delta)}{2 (s+2 \Delta-\tau -2 m ) (u-\tau -2 m )}\right)
    \,,
\end{multline}
where we recall that $s+t+u=2\Delta$. Let's consider the contributions in the direct channel expansion. In Mellin space \eqref{Mellinrepresentation}, these are encoded in the poles of the Mellin variable $t$ that lie to the right of the integration contour.
From the Mellin exchange amplitude $\mathcal{E}_{\tau,\ell}(s,t)$ we have only the single poles at $t=\tau+2m$, which are:
\begin{equation}\label{STpoledirect}
       \frac12\frac{\mathcal{Q}_{\tau,\ell|m}(s)+\mathcal{Q}_{\tau,\ell|m}(-s-\tau-2m)}{t-\tau-2m}.
\end{equation}
These originate from the principal part of the Laurent expansion of the first two terms in the summand of \eqref{crossExch} and encode the primary $\left(m=0\right)$ and descendent $\left(m>0\right)$ operator contributions dual to the spin-$\ell$ field in AdS exchanged in the ${\sf s}$-channel.

On top of the above single-trace contribution, the function $\rho(s,t)$ in \eqref{Mellinrepresentation} has an infinite number of double-poles at $t=2\Delta+2n$, $n=0,1,2,3,...$, corresponding to contributions from Regge-trajectories of double-twist operators of spin $\ell^\prime=0,1,2,3,\ldots$. These contributions come in two types:

\begin{enumerate}
    \item \underline{Contributions analytic in spin $\ell^\prime$} originate from the single poles in the remaining Mellin variable $s$ in the Mellin exchange amplitude $\mathcal{E}_{\tau,\ell}(s,t)$, which are the two families: 
    \begin{subequations}\label{spoles}
    \begin{align}
        s&=-2\Delta+\tau-2m,\\
        s&=-t+2\Delta-\tau-2m,
    \end{align}
    \end{subequations}
    corresponding to the exchange of twist $\tau$ primary operators (+descendants) in the crossed channels. The corresponding terms in the Mellin amplitude \eqref{crossExch} read
    \begin{multline}
          \frac{(t-4 \Delta+2\tau +4 m)}{2 (s+2 \Delta -\tau-2 m ) (-s-t+2 \Delta-\tau-2m )}\\ \times \left[\mathcal{Q}_{\tau,\ell|m}(t-2 \Delta )+\mathcal{Q}_{\tau,\ell|m}(2 \Delta -2 m-t-\tau )\right],
    \end{multline}
    which are the sum of the principle parts of the Laurent expansions around the poles \eqref{spoles} in $s$. These terms are suppressed in the Regge limit $s \to \infty$ and $t$ fixed since $s$ only appears in the denominator. The corresponding OPE data is thus analytic in spin down to spin $\ell^\prime=0$. 
    
    \item \underline{Terms non-analytic in spin $\ell^\prime$} correspond to polynomials in the Mellin variable $s$. These can be written down explicitly and can be divided into two types: A term that is polynomial in $s$ but singular in $t$,
    \begin{subequations}
    \begin{align}
        \mathfrak{p}_{\tau,\ell|m}(s|t)=\frac12\frac{\mathcal{Q}_{\tau,\ell|m}(s)+\mathcal{Q}_{\tau,\ell|m}(-2 m-s-\tau )}{t-2 m- \tau }\,,\label{p_pol}
    \end{align}
    \end{subequations}
    and a term that is polynomial in both $s$ and $t$, 
    \begin{multline}\label{qpol_cross}
        \mathfrak{q}_{\tau,\ell|m}(s|t)=\frac{1}{2} \Bigg(\tfrac{\frac{(2 \Delta -4 m+s+t-2 \tau )\,\mathcal{Q}_{\tau,\ell|m}(-s-t)}{-2 \Delta +2 m-s+\tau }+\mathcal{Q}_{\tau,\ell|m}(-2 m-s-\tau )}{2 m-t+\tau }\\+\tfrac{(2 \Delta -2 m+s-\tau )\mathcal{Q}_{\tau,\ell|m}(s)+(t-4 \Delta +2 \tau +4 m) \mathcal{Q}_{\tau,\ell|m}(2 \Delta -2 m-t-\tau )}{(-2 \Delta +2 m-s+\tau ) (-2 \Delta +2 m+s+t+\tau )}\Bigg).
    \end{multline}
    We remind the reader that the above expressions are assumed to be evaluated on one of the double-trace poles $t=2\Delta+2n$.
    Notice that only $\mathfrak{p}_{\tau,\ell|m}(s|t)$ is a polynomial of $\text{degree $\ell$}$ in $s$ while $\mathfrak{q}_{\tau,\ell|m}(s|t)$ is a polynomial in $s$ of degree $\ell-1$.
\end{enumerate}

The contact term ambiguity arises because the full crossing symmetric exchange amplitude \eqref{crossExch} is unbounded in the Regge limit, where the leading behaviour for $s \to \infty$ and $t$ fixed is given by the degree $\ell$ polynomials $\mathfrak{p}_{\tau,\ell|m}(s|t)$ which are determined by the residues of the single-trace poles \eqref{STpoledirect} in the direct channel. One is therefore allowed to add crossing symmetric polynomial terms $p_{\ell-1}\left(s,t\right)$ of degree $\ell-1$ or less to $\mathfrak{q}_{\tau,\ell|m}(s|t)$, i.e. $\mathfrak{q}_{\tau,\ell|m}(s|t)\sim \mathfrak{q}_{\tau,\ell|m}(s|t)+p_{\ell-1}(s,t)$, which is discussed in detail in Appendix \ref{ContactAmb}.

As discussed above, a minimal prescription to fix the contact term ambiguity is to require that:
\begin{equation}\label{idextcond}
    \mathfrak{q}_{\tau,\ell|m}(s|t) \in \ker \mathfrak{C}\circ\mathcal{S}.
\end{equation}
This is precisely the case for the $\mathfrak{q}_{\tau,\ell|m}(s,t)$ in \eqref{crossExch} corresponding to our initial choice: ${p_{\ell-1}\left(s,t\right)\equiv 0}$. Any other choice of contact term would violate the condition \eqref{idextcond}, which makes clear that it is a universal contribution to exchange amplitudes. With this prescription, the Polyakov block for identical external legs is thus given unambiguously by \eqref{crossExch}. It is immediate to check that this is identical to the Polyakov block \eqref{PBidleg} singled out by the requirement that the cyclic exchange amplitudes \eqref{cyclic_exch} are suppressed in the Regge limit, so that the two prescriptions are equivalent in the case of identical external legs. The prescription defined at the level of the cyclic exchange amplitudes in section \ref{sec::problemandresolution} is however more general, as it also applies in the case where the external legs are unequal. The cyclicly ordered Polyakov blocks \eqref{cyclic_exch_fin} should therefore be regarded as the minimal building blocks of crossing symmetric solutions.

In the following section \ref{sec::CBdecomp} we shall determine the conformal block expansion of the cyclicly ordered Polyakov blocks \eqref{cyclic_exch_fin} in the direct channel. To this end, for each cyclic exchange amplitude \eqref{cyclic_exch} it is convenient to separate the pole and polynomial parts in $s$ as we did for the exchange amplitude \eqref{crossExch} above which, respectively, identify the terms which generate analytic and non-analytic contributions in spin $\ell^\prime$ of the double-twist operators. Starting with $\mathcal{E}^{(1234)}_{\tau,\ell}(s,t)$, and writing
\begin{equation}
\mathcal{E}^{(1234)}_{\tau,\ell}(s,t)=\sum^\infty_{m=0}   \mathcal{E}^{(1234)}_{\tau,\ell|m}(s,t),
\end{equation}
we obtain:
\begin{shaded}
\begin{multline}\label{s-expansion1234}
    \mathcal{E}^{(1234)}_{\tau,\ell|m}(s,t)=\underbrace{\frac{\mathcal{Q}_{\tau,\ell|m}(-t+2\Delta-\tau-2m)}{-s-t+2\Delta-\tau-2m}}_{\mathfrak{A}^{(\sf{u})}_{\tau,\ell|m}(s|t)}+\underbrace{\frac{\mathcal{Q}_{\tau,\ell|m}(s)}{t-\tau-2m}}_{\mathfrak{p}^{(1234)}_{\tau,\ell|m}(s|t)\ \text{, universal polynomial in $s$ of degree $\ell$}}\\+\underbrace{\mathfrak{q}^{(1234)}_{\tau,\ell|m}(s|t)}_{\text{part subject to contact term ambiguity}}\,,
\end{multline}
\end{shaded}
\noindent with
\begin{equation}
    \mathfrak{q}^{(1234)}_{\tau,\ell|m}(s|t)\equiv \frac{\mathcal{Q}_{\tau,\ell|m}(2 \Delta -2 m-t-\tau )-\mathcal{Q}_{\tau,\ell|m}(s)}{-2 \Delta +2 m+s+t+\tau }+p_{\ell-1}(t,2 \Delta -s-t)\,,
\end{equation}
is a polynomial of degree $\ell-1$ in the variable $s$ (and $t$) and $\mathfrak{A}^{(\sf{u})}_{\tau,\ell|m}(s|t)$ is instead the principal part in the variable $s$ of the cyclic amplitude $\mathcal{E}^{(1234)}_{\tau,\ell}(s,t)$. For $\mathcal{E}^{(1342)}_{\tau,\ell}(s,t)$ one gets:
\begin{shaded}
\begin{multline}
    \mathcal{E}^{(1342)}_{\tau,\ell|m}(s,t)=\underbrace{\frac{\mathcal{Q}_{\tau,\ell|m}(-t+2\Delta-\tau-2m)}{s+2\Delta-\tau-2m}}_{(-1)^\ell\mathfrak{A}^{(\sf{t})}_{\tau,\ell|m}(s|t)}+\underbrace{\frac{\mathcal{Q}_{\tau,\ell|m}(-s-\tau-2m)}{t-\tau-2m}}_{\mathfrak{p}^{(1342)}_{\tau,\ell|m}(s|t)\ \text{, universal polynomial in $s$ of degree $\ell$}}\\+\underbrace{\mathfrak{q}^{(1342)}_{\tau,\ell|m}(s|t)}_{\text{part subject to contact term ambiguity}}\,,
\end{multline}
\end{shaded}
\noindent where now the degree $\ell-1$ polynomial in $s$ reads:
\begin{multline}
    \mathfrak{q}^{1342}_{\tau,\ell|m}(s|t)\equiv \mathcal{Q}_{\tau,\ell|m}(-s-t) \left(\frac{1}{2 \Delta -2 m+s-\tau }+\frac{1}{-2 m+t-\tau }\right)+\frac{\mathcal{Q}_{\tau,\ell|m}(2 \Delta -2 m-t-\tau )}{-2 \Delta +2 m-s+\tau }\\+\frac{\mathcal{Q}_{\tau,\ell|m}(-2 m-s-\tau )}{2 m-t+\tau }+p_{\ell-1}(2 \Delta +s,t)\,,
\end{multline}
and we have also introduced the $\sf{t}$-channel principal part $\mathfrak{A}^{(\sf{t})}_{\tau,\ell|m}(s|t)$ in the variable $s$. Finally, for $\mathcal{E}^{(1423)}_{\tau,\ell}$ we have:
\begin{shaded}
\begin{multline}
    \mathcal{E}^{(1423)}_{\tau,\ell|m}(s,t)=\underbrace{\frac{\mathcal{Q}_{\tau,\ell|m}(t-2\Delta)}{s+2\Delta-\tau-2m}}_{\mathfrak{A}^{(\sf{t})}_{\tau,\ell|m}(s|t)}+\underbrace{\frac{\mathcal{Q}_{\tau,\ell|m}(t-2\Delta)}{-s-t+2\Delta-\tau-2m}}_{(-1)^\ell\,\mathfrak{A}^{(\sf{u})}_{\tau,\ell|m}(s|t)}+\underbrace{\mathfrak{q}^{1342}_{\ell,m}(s|t)}_{\text{part subject to contact term ambiguity}}\,,
\end{multline}
\end{shaded}
\noindent where now the term of degree $\ell$ in the Mellin variable $s$ is absent while the degree $\ell-1$ polynomial in $s$ is entirely given by the polynomial $p_{\ell-1}(X,Y)$ parametrising the contact term ambiguity:
\begin{align}
    \mathfrak{q}^{(1423)}_{\tau,\ell|m}(s|t)\equiv p_{\ell-1}(2 \Delta -s-t,2 \Delta +s)\,.
\end{align}
We give a few comments about the above decompositions below:
\begin{itemize}
    \item The kinematic polynomials $\mathcal{Q}_{\tau,\ell|m}(s)$ satisfy the following identity:
    \begin{align}\label{idQ}
       \mathcal{Q}_{\tau,\ell|m}(s)=(-1)^\ell \mathcal{Q}_{\tau,\ell|m}(-s-\tau-2m)\,.
    \end{align}
    This implies that there are only two independent meromorphic terms in $s$ that appear in the cyclic amplitudes. These are associated to the principal parts $\mathfrak{A}^{(\sf{t})}_{\tau,\ell|m}(s|t)$ and $\mathfrak{A}^{(\sf{u})}_{\tau,\ell|m}(s|t)$ in the Laurent expansion of the $\sf{t}$ and $\sf{u}$ channel exchanges in $s$, respectively. In the case of equal external scaling dimensions dimensions $\Delta$, both $\sf{t}$ and $\sf{u}$ channel contributions give OPE data in the direct channel which are equal up to a sign $(-1)^\ell$ (see e.g. \cite{Sleight:2018ryu}).
    \item The identity \eqref{idQ} moreover implies that the universal degree-$\ell$ polynomials in $s$, giving the leading behaviour of $\mathcal{E}^{(1234)}_{\tau,\ell|m}(s,t)$ and $\mathcal{E}^{(1342)}_{\tau,\ell|m}(s,t)$ in the Regge limit $s \to \infty$ and $t$ fixed, are equal up to a sign, 
        \begin{equation}
    \mathfrak{p}^{(1234)}_{\tau,\ell|m}(s|t)=\left(-1\right)^\ell \mathfrak{p}^{(1342)}_{\tau,\ell|m}(s|t).
    \end{equation}

    \item In $\mathcal{E}^{(1423)}_{\tau,\ell|m}(s,t)$ the only polynomial contribution in $s$ is entirely proportional to $p_{\ell-1}(X,Y)$. This implies that the contact term ambiguity $p_{\ell-1}(X,Y)$ parameterises the non-analyticity in spin of the conformal block decomposition of $\mathcal{E}^{(1423)}_{\tau,\ell|m}(s,t)$ in the direct-channel. Fixing $p_{\ell-1}(X,Y)=0$ is equivalent to requiring that the conformal block decomposition of $\mathcal{E}^{(1423)}_{\tau,\ell|m}(s,t)$ in the direct-channel is analytic in spin up to spin zero. In fact, with the prescription \eqref{cyclic_exch_fin}, for each cyclic amplitude there is always one channel in which the corresponding OPE decomposition is analytic up to spin $\ell=0$!
\end{itemize}

\section{Conformal block decomposition of Polyakov Blocks}
\label{sec::CBdecomp}

Having uniquely fixed the cyclic Polyakov blocks, in this section we determine their conformal block decomposition in the direct channel, extracting explicit expressions for the OPE data. In section \ref{subsec::abfunc} we discuss the relation between the cyclic Polyakov blocks \eqref{cyclic_exch_fin} and analytic bootstrap functionals.

\subsection{Conformal block decomposition}

The conformal block decomposition of the cyclicly ordered Polyakov blocks in the direct channel takes the form:
\begin{subequations}\label{CBexppos}
\begin{align}
    \mathcal{E}_{\tau,\ell}^{\bullet}(u,v)&=g_{\tau,\ell}(u,v)+\sum_{n,\ell^\prime=0}^\infty a^{\bullet}_{n,\ell^\prime|\tau,\ell}\, g_{\tau_1+\tau_2+2n,\ell^\prime}(u,v)+\sum_{n,\ell^\prime=0}^\infty b^{\bullet}_{n,\ell^\prime|\tau,\ell}\, g_{\tau_3+\tau_4+2n,\ell^\prime}(u,v)\,,\\
    \mathcal{E}^{(1342)}_{\tau,\ell}(u,v)&=\sum_{n,\ell^\prime=0}^\infty a^{\bullet}_{n,\ell^\prime|\tau,\ell}\, g_{\tau_1+\tau_2+2n,\ell^\prime}(u,v)+\sum_{n,\ell^\prime=0}^\infty b^{\bullet}_{n,\ell^\prime|\tau,\ell}\, g_{\tau_3+\tau_4+2n,\ell^\prime}(u,v),\,
\end{align}
\end{subequations}
\noindent where $\bullet=1234$ or $1423$, which each receive a contribution from the conformal block of twist $\tau$ and spin-$\ell$ generated by the single poles at $t=\tau+2m$ in their Mellin representation. The latter are not present in  $\mathcal{E}^{(1342)}_{\tau,\ell}\left(s,t\right)$. All cyclic Polyakov blocks have contributions from families of double-twist operators, which are generated by the two families of poles at $t=\tau_1+\tau_1+2n$ and $t=\tau_3+\tau_4+2n$ in the Mellin measure $\rho_{\tau_i}(s,t)$ given explicitly in \eqref{rhouequal}.  For the moment we have assumed that the external twists are unequal. When the external twists are equal the two families of double-twist poles coincide, generating anomalous dimensions, which can be obtained by carefully taking the limit of the above decomposition as we shall see below.

 The coefficients $a^{\bullet}_{n,\ell^\prime|\tau,\ell}$ and $b^{\bullet}_{n,\ell^\prime|\tau,\ell}$ can be extracted systematically from the expressions \eqref{cyclic_exch_fin} for the cyclic Polyakov blocks in Mellin space, using that the kinematic polynomials $\mathcal{Q}_{\tau,\ell|m}(s)$ with $m=0$ are orthogonal \cite{Costa:2012cb}, being proportional to so-called Continuous Hahn polynomials $Q_\ell^{(\tau,\tau+\tau_1-\tau_2-\tau_3+\tau_4,-\tau_1+\tau_2,\tau_3-\tau_4)}(s)$ which are orthogonal with respect to the measure \eqref{rhotilde} (see appendix \ref{Hahn}). This is detailed in \cite{Sleight:2018epi}. In particular, one can always reduce the problem of extracting conformal block coefficients in a given channel to that of extracting the coefficient of the leading twist contribution, which can be obtained through the inversion formula \cite{Sleight:2018epi}:
\begin{align}
    a_{\tau_{min},\ell|\tau,\ell}=\frac{(-1)^\ell}{\ell!} \int_{-i\infty}^{+i\infty}\frac{ds}{4\pi i}\,\tilde{\rho}_{\tau_i}(s,\tau)\widehat{\mathcal{M}}(s,\tau_{min})\,Q_\ell^{(\tau_{min},\tau_{min}+\tau_1-\tau_2-\tau_3+\tau_4,-\tau_1+\tau_2,\tau_3-\tau_4)}(s)\,,
\end{align}
where $\tau_{min}$ is the lowest twist appearing in the direct channel expansion of a correlator with Mellin amplitude $\mathcal{M}(s,t)$,
\begin{align}
    \tilde{\rho}(s,\tau_{min})\widehat{\mathcal{M}}(s,\tau_{min})=-\tfrac12\text{Res}_{t=\tau_{min}}\left[\rho_{\{\tau_i\}}(s,t)\mathcal{M}(s,t)\right]\,,
\end{align} 
and measure
\begin{align}\label{rhotilde}
    \tilde{\rho}_{\{\tau_i\}}(s,t)&=\Gamma\left(\tfrac{s+t}2\right)\Gamma\left(\tfrac{s+t+\tau_1-\tau_2-\tau_3+\tau_4}2\right)\Gamma\left(\tfrac{-s-\tau_1+\tau_2}2\right)\Gamma\left(\tfrac{-s+\tau_3-\tau_4}2\right)\,.
\end{align} 
 It is convenient to introduce the following one parameter family of inner products:
\begin{align}
    \left\langle f(s)\Big|g(s)\right\rangle_t=\int_{-i\infty}^{+i\infty}\frac{ds}{4\pi i}\,\tilde{\rho}_{\tau_i}(s,t)f(s)g(s)\,,
\end{align}
defined on functions of a single Mellin variable $s$. In this way the coefficients $a_{n,\ell}$ and $b_{n,\ell}$ in the conformal block expansions \eqref{CBexppos} can be expressed as the following functional action:
\begin{equation}\label{func}
    \omega_{n,\ell^\prime}^{(i)}\left[\mathcal{E}_{\tau,\ell}\right]\equiv\left\langle{{}^{(n)}\mathcal{E}_{\tau,\ell}}(s,\tau_{min}^i+2n)\Big| \frac{(-1)^{\ell^\prime}}{\ell^\prime!}\,Q_{\ell^\prime}^{(\tau^{i}_{min}+2n,\tau^i_{min}+2n+\tau_1-\tau_2-\tau_3+\tau_4,-\tau_1+\tau_2,\tau_3-\tau_4)}\right\rangle_{\tau_{min}^i+2n}\,,
\end{equation}
where the superscript ${}^{(n)}$ indicates the projection operation:
\begin{equation}\label{twistproj}
   {}^{(n)}\mathcal{E}_{\tau,\ell}=\left(\widehat{ \mathcal{T}}_{\tau_{min}^i}\right)^n[\mathcal{E}_{\tau,\ell}]\,,
\end{equation}
introduced in eq.~(\tcb{5.30}) of \cite{Sleight:2018epi}, which projects away all contributions from operators with twist $\tau_{\min}^i$ up to $\tau_{\min}^i+2(n-1)$, while the superscript ${}^i$ labels the minimum twist which can be either $\tau_1+\tau_2$ or $\tau_3+\tau_4$. Equipped with the above definitions, the conformal block decompositions \eqref{CBexppos} are given by:
\begin{subequations}
\begin{align}
    \mathcal{E}^{\bullet}_{\tau,\ell}(u,v)&=g_{\tau,\ell}(u,v)+\sum_{n,\ell^\prime=0}^\infty \omega_{n,\ell^\prime}^{(1)}\left[\mathcal{E}_{\tau,\ell}^\bullet\right]\, g_{\tau_1+\tau_2+2n,\ell^\prime}(u,v)+\sum_{n,\ell^\prime=0}^\infty \omega_{n,\ell^\prime}^{(2)}\left[\mathcal{E}_{\tau,\ell}^\bullet\right]\, g_{\tau_3+\tau_4+2n,\ell^\prime}(u,v)\,,\\
    \mathcal{E}^{(1342)}_{\tau,\ell}(u,v)&=\sum_{n,\ell^\prime=0}^\infty \omega_{n,\ell^\prime}^{(1)}\left[\mathcal{E}_{\tau,\ell}^{(1342)}\right]\, g_{\tau_1+\tau_2+2n,\ell^\prime}(u,v)+\sum_{n,\ell^\prime=0}^\infty \omega_{n,\ell^\prime}^{(2)}\left[\mathcal{E}_{\tau,\ell}^{(1342)}\right]\, g_{\tau_3+\tau_4+2n,\ell^\prime}(u,v).
\end{align}
\end{subequations}
and what remains is to evaluate the inner products \eqref{func}. Without loss of generality we shall focus on the cyclic exchange amplitude with $\bullet=(1234)$.

We shall be particularly interested in the case where the scaling dimensions of the external operators are equal, $\tau_i=\Delta$, where the double-twist operators appearing in the conformal block decomposition \eqref{CBexppos} receive anomalous dimensions $\gamma_{n,\ell^\prime|\tau,\ell}$. These are given by the coefficient of the derivative of the corresponding double-twist conformal block with respect to the twist, which appear in the limit where the external scaling dimensions coincide. This can be seen by taking $\tau_{i=1,2,3}=\Delta$ with $\tau_4=\Delta+\epsilon$ in \eqref{CBexppos} and expanding in $\epsilon$. Both $a_{n,\ell^\prime|\tau,\ell}(\epsilon)$ and $b_{n,\ell^\prime|\tau,\ell}(\epsilon)$ have simple poles in $\epsilon$ and their expansion around $\epsilon=0$ takes the form:\footnote{This equation follows directly from the structure of the Mellin integral so that in the limit $\epsilon\to0$ two single poles collide into a double-pole. This implies that the single terms $a_{n,\ell^\prime|\tau,\ell}(\epsilon)$ and $b_{n,\ell^\prime|\tau,\ell}(\epsilon)$ have singularities but their combination does not!} 
\begin{subequations}
\begin{align}
a_{n,\ell^\prime|\tau,\ell}(\epsilon)&\sim\frac{a^{(0)}_{n,\ell^\prime}\gamma_{n,\ell^\prime|\tau,\ell}}{2\epsilon}+\bar{a}_{n,\ell^\prime|\tau,\ell}+O(\epsilon)\,,\\
b_{n,\ell^\prime|\tau,\ell}(\epsilon)&\sim-\frac{a^{(0)}_{n,\ell^\prime}\gamma_{n,\ell^\prime|\tau,\ell}}{2\epsilon}+\bar{b}_{n,\ell^\prime|\tau,\ell}+O(\epsilon)\,,
\end{align}
\end{subequations}
where $a^{(0)}_{n,\ell^\prime}$ is the Mean Field Theory OPE coefficients \cite{Dolan:2000ut,Heemskerk:2009pn}
\begin{equation}\label{MFTOPE}
    a^{(0)}_{n,\ell^\prime}=\frac{2^{\ell^\prime} (-1)^n (\Delta )_n^2 \left(-\frac{d}{2}+\Delta +1\right)_n^2 (n+\Delta )_{\ell^\prime}^2}{\ell^\prime! n! \left(\frac{d}{2}+\ell^\prime\right)_n (d-2 n-2 \Delta )_n (\ell^\prime+2 n+2 \Delta -1)_{\ell^\prime} \left(-\frac{d}{2}+\ell^\prime+n+2 \Delta \right)_n},
\end{equation}
so that for equal external scaling dimensions $\Delta$ we have
\begin{multline}\label{exchangedec2}
\mathcal{E}_{\tau,\ell}(u,v)=g_{\tau,\ell}(u,v)+\sum_{n,\ell^\prime=0}^\infty (\bar{a}_{n,\ell^\prime|\tau,\ell}+\bar{b}_{n,\ell^\prime|\tau,\ell})\, g_{2\Delta+2n,\ell^\prime}(u,v)\\-\sum_{n,\ell^\prime=0}^\infty \frac{a^{(0)}_{n,\ell^\prime}\gamma_{n,\ell^\prime|\tau,\ell}}{2}\pl g_{2\Delta+2n,\ell^\prime}(u,v)\,,
\end{multline}
where $\pl g\equiv\pl_\tau g_{\tau,\ell}$ represents the additional basis element needed to have a well-defined expansion in this case. As before the corresponding coefficients can be expressed as functionals acting on the Mellin representation of the cyclic exchange amplitude,
\begin{align}
\omega_{n,\ell^\prime}[\mathcal{E}_{\tau,\ell}]&=\bar{a}_{n,\ell^\prime|\tau,\ell}+\bar{b}_{n,\ell^\prime|\tau,\ell}\,,& \omega_{n,\ell^\prime}^\pl[\mathcal{E}_{\tau,\ell}]&=-\frac{a^{(0)}_{n,\ell^\prime}\gamma_{n,\ell^\prime|\tau,\ell}}{2}\,,
\end{align}
which are inherited from those \eqref{func} for generic external scaling dimensions in the limit $\epsilon\to0$.

There are three different types of contributions to the double-twist OPE data \eqref{func} in the direct channel expansion of the cyclic Polyakov block $\mathcal{E}^{(1234)}_{\tau,\ell}$, which can be recognised from the expression \eqref{s-expansion1234}:

\begin{enumerate}
    \item \underline{Analytic in spin OPE data} is generated by the single poles in the Mellin variable $s$ and is given by
    \begin{align}\label{analspin_OPE}
\omega^{(i)}_{n,\ell^\prime}&\left[\sum^\infty_{m=0} \mathfrak{A}^{(\sf{u})}_{\tau,\ell|m}(s|\tau_{min}^i+2n)\right]\,.
\end{align}
These OPE data, which is analytic in spin down to spin $\ell^\prime=0$, was extracted in \cite{Sleight:2018ryu} using the approach outlined above, where they were expressed in terms of Wilson functions,\footnote{Wilson functions can be expressed in various convenient ways, for instance: as a ``very well poised” hypergeometric function ${}_{7}F_{6}$, a combination of 1-balanced hypergeometric functions ${}_4F_3$, or in terms of integrated products of Gauss hypergeometric functions ${}_2F_1$. In the examples below we shall employ the latter representation.} which ensures that analyticity in spin is manifest.\footnote{As a word of caution, this is contrary to some other expressions available for some of these data in the literature.}
\item \underline{Non-analytic in spin contributions for spin $\ell^\prime=\ell$} come from the universal polynomial terms $\mathfrak{p}_{\tau,\ell|m}^{(1234)}\left(s|t\right)$ of degree $\ell$ in $s$, 
    \begin{align}\label{lpeqlfsup}
\omega^{(i)}_{n,\ell^\prime=\ell}&\left[\sum^\infty_{m=0} \mathfrak{p}^{(1234)}_{\ell,m}(s|2\Delta+2n)\right]\,.
\end{align}
In appendix \ref{pterm} we show that, taking the limit of equal external scaling dimensions $\tau_i=\Delta$,  
\begin{align}\label{ContactGen}
    \omega_{n,\ell^\prime=\ell}^{\pl}\left[\sum^\infty_{m=0} \mathfrak{p}^{(1234)}_{\ell,m}(s|2\Delta+2n)\right]&=-\frac12a^{(0)}_{n,\ell}\gamma^{\text{n.-a.}}_{n,\ell|\tau,\ell},
\end{align}
with
\begin{shaded}
\begin{multline}\label{gammaEll}
    \gamma^{\text{n.-a.}}_{n,\ell|\tau,\ell}=\gamma^{\text{n.-a.}}_{0,0|\tau,0}\,\frac{(\tau -2 \Delta ) (d-2 \Delta -\tau )}{(2 \Delta +2 n-\tau ) (-d+2 \Delta +2 \ell+2 n+\tau )}\\\times\,\tfrac{2^{-\ell} \ell! \left(\frac{\tau +1}{2}\right)_\ell (\ell+\tau -1)_\ell \left(\frac{d}{2}\right)_{\ell+n} \left(\frac{d-2 (\tau +1)}{2}-\ell+1\right)_\ell \left(\frac{d-2 \Delta -1}{2}-n+1\right)_n (\Delta )_{\ell+n} \left(\frac{d-4 \Delta}{2}-\ell-n+1\right)_{\ell+n}}{n! \left(\frac{d}{2}\right)_\ell \left(\frac{\tau }{2}\right)_\ell^3 \left(\frac{d-2 \Delta -2}{2}-n+1\right)_n (d-n-2 \Delta )_n \left(\frac{2 \Delta +1}{2}\right)_{\ell+n} \left(\frac{2 \Delta +\tau-d}{2}\right)_\ell^2}\,,
\end{multline}
\end{shaded}
where 
\begin{shaded}
\begin{equation}\label{gammaEll00}
    \gamma^{\text{n.-a.}}_{0,0|\tau,0}=\frac{2^{-2 \Delta +\tau +1} \Gamma (\Delta )^3 \Gamma \left(\frac{\tau +1}{2}\right) \Gamma \left(2 \Delta -\frac{d}{2}\right) \Gamma \left(-\frac{d}{2}+\tau +1\right)}{\Gamma \left(\Delta +\frac{1}{2}\right) \Gamma \left(\frac{\tau }{2}\right)^3 \Gamma \left(\Delta -\frac{\tau }{2}\right) \Gamma \left(\Delta -\frac{\tau }{2}+1\right) \Gamma \left(\Delta -\frac{d-\tau }{2}\right) \Gamma \left(\Delta -\frac{d-\tau}{2}+1\right)}\,.
\end{equation}
\end{shaded}
To the best of our knowledge, the above result is new and reduces to the result of \cite{Alday:2017gde} obtained by a case by case study tuned to $\tau=2$ and $d=4$.  One can similarly extract the corresponding OPE coefficients $\bar{a}_{n,\ell|\tau,\ell}+\bar{b}_{n,\ell|\tau,\ell}$ and we give some examples below.

\item \underline{Non-analytic in spin contributions for spin $\ell^\prime<\ell$} come not only from $\mathfrak{p}_{\tau,\ell|m}^{(1234)}\left(s|t\right)$ but also the polynomial $\mathfrak{q}_{\tau,\ell|m}^{(1234)}(s|t)$ of degree less than $\ell$ in $s$. Such contributions are thus given by the inner product:
\begin{equation}\label{lpleqlfsup}
    \omega^{(i)}_{n,\ell^\prime}\left[\sum_m\mathfrak{p}_{\tau,\ell|m}^{(1234)}(s|2\Delta+2n)+\mathfrak{q}_{\tau,\ell|m}^{(1234)}(s|2\Delta+2n)\right].
\end{equation}
and in the following we give some examples. 
\end{enumerate}

\paragraph{Scalar exchange $\ell=0$.} This is the simplest case, in which  $\mathfrak{q}_{\tau,0|m}^{(\bullet)}(s|t)\equiv0$. The cyclic Polyakov blocks for equal external scaling dimensions $\Delta$ read 
\begin{subequations}
\begin{align}
\mathcal{E}^{(1234)}_{\tau,0|m}&=\frac{2\mathfrak{c}^{(0)}_m}{-2 \Delta +2 m+s+t+\tau }+\frac{2\mathfrak{c}^{(0)}_m}{2 m-t+\tau }\,,\\
\mathcal{E}^{1342}_{\tau,0|m}&=\frac{2\mathfrak{c}^{(0)}_m}{-2 \Delta +2 m+s+t+\tau }+\frac{2\mathfrak{c}^{(0)}_m}{-2 \Delta +2 m-s+\tau }\,,\\
\mathcal{E}^{1423}_{\tau,0|m}&=\frac{2\mathfrak{c}^{(0)}_m}{-2 \Delta +2 m-s+\tau }+\frac{2\mathfrak{c}^{(0)}_m}{+2 m-t+\tau }\,,
\end{align}
\end{subequations}
where the coefficients $\mathfrak{c}^{(0)}_m$ are defined in \eqref{clm}. In the following we focus without loss of generality on extracting the OPE data of $\mathcal{E}^{1234}_{\tau,0}$ in the direct channel. 

Let us first consider contributions \eqref{analspin_OPE} to the OPE data that are analytic in spin. For the anomalous dimensions of the double-twist operators, the corresponding inner product was already evaluated in \cite{Sleight:2018ryu} and for those operators with leading twist $2\Delta$ (i.e. $n=0$) it reads:
\begin{align}\label{scalar_anal}
  -\frac12 a^{(0)}_{0,\ell^\prime}\gamma^{\text{anal.}}_{0,\ell^\prime|\tau,0}&= {\omega}^{\pl}_{0,\ell^\prime}\left[\sum^\infty_{m=0}\mathfrak{A}^{(\sf{u})}_{\tau,\ell=0|m}(s,2\Delta)\right]\\ \nonumber
  &=\frac{\Gamma (\Delta )^2 \Gamma (\tau ) \Gamma \left(-\frac{d}{2}+\ell^\prime+2 \Delta \right)}{\Gamma \left(\frac{\tau }{2}\right)^2 \Gamma \left(-\frac{d}{2}+\tau +1\right) \Gamma \left(\Delta -\frac{\tau }{2}\right)^2 \Gamma (\ell^\prime+2 \Delta -1)}\\ \nonumber
  & \hspace*{3cm}\times\int_{0}^1 dy\,(1-y)^{\ell^\prime} y^{\tau -\frac{d}{2}} \, _2F_1\left(\begin{matrix}\tfrac{\tau+2-d}{2},\tfrac{\tau-2\Delta+2}{2}\\\tau-\tfrac{d}{2} +1\end{matrix};y\right)^2\,,
\end{align}
where we employed the representation of the Wilson function given by an integrated product of two Gauss hypergeometric functions (see equation (2.29) of \cite{Sleight:2018ryu}). This expression is analytic in spin down to $\ell^\prime=0$. 

\begin{figure}
    \centering
    \includegraphics[width=0.6\textwidth]{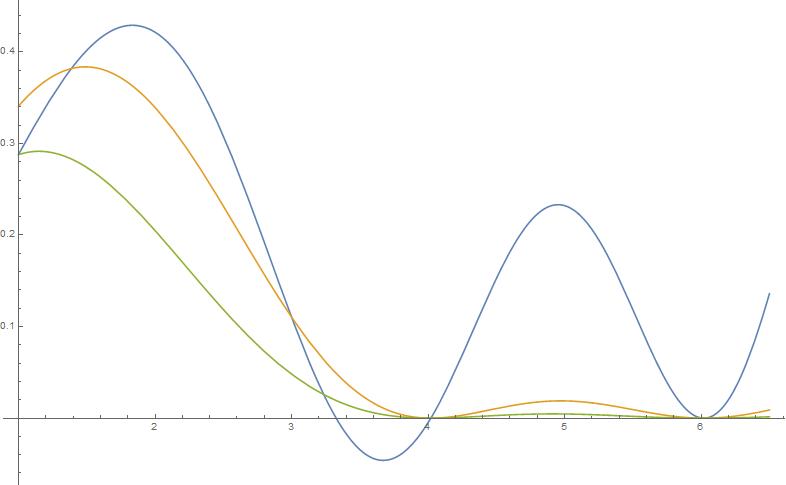}
    \caption{Plot of $\gamma_{0,\ell^\prime|\tau,0}=\gamma^{\text{anal.}}_{0,\ell^\prime|\tau,1}+\gamma^{\text{n.-a.}}_{0,\ell^\prime|\tau,0}$ for $d=3$, $\Delta=2$ and varying $\tau$ along the $x$ axis. In blue we have $\ell^\prime=0$, in orange $\ell^\prime=1$ and in green $\ell^\prime=2$. We see that for $\ell>0$, where there are only the analytic in spin contributions \eqref{scalar_anal}, $\gamma_{0,\ell^\prime|\tau,0}$ is always positive and has double zeros at double-twist values of $\tau$. For $\ell^\prime=0$ there is also a non-analytic contribution \eqref{gammaEll}, so that the full anomalous dimension $\gamma_{0,\ell^\prime|\tau,0}$ also has single zeros at some double-twist values of $\tau$.}
    \label{fig:graph1}
\end{figure}

 In a similar way one can obtain the corresponding double-twist OPE coefficients in \eqref{exchangedec2}:
\begin{align}\label{OPEomega}
\bar{a}^{\text{anal.}}_{0,\ell^\prime|\tau,0}+\bar{b}^{\text{anal.}}_{0,\ell^\prime|\tau,0}&=\omega_{0,\ell^\prime}\left[\sum^\infty_{m=0}\mathfrak{A}^{(\sf{u})}_{\tau,\ell=0|m}(s,2\Delta)\right]\\ \nonumber
&=\frac{2^{-2 \Delta -\ell^\prime+\tau +1}\Gamma \left(\frac{\tau +1}{2}\right) \Gamma (\ell^\prime+\Delta )^3}{\ell^\prime!\, \Gamma \left(\frac{\tau }{2}\right) \Gamma \left(\frac{d}{2}+\ell^\prime\right) \Gamma \left(-\frac{d}{2}+\tau +1\right) \Gamma \left(\Delta -\frac{\tau }{2}\right)^2 \Gamma \left(\ell^\prime+\Delta -\frac{1}{2}\right)}\\\nonumber
    &\times\int_0^1 dy\,(y-1)^{\ell^\prime} y^{\tau -\frac{d}{2}} \, _2F_1\left(\begin{matrix}\tfrac{\tau-2 \Delta +2}{2},\tfrac{\tau-2 \Delta  +2}{2}\\-\tfrac{d}{2}+\tau +1\end{matrix};y\right) \\ & \hspace*{1cm} \times \Bigg[2\, \Xi_{\ell^\prime}\ \, _2F_1\left(\begin{matrix}\tfrac{\tau-d +2}{2},\tfrac{\tau-d+2}{2}\\\tau-\tfrac{d}{2} +1\end{matrix};y\right)-2\pl_x\ {}_2F_1\left(\begin{matrix}\tfrac{\tau-d +2}{2},x+\tfrac{\tau-d+2}{2}\\\tau-\tfrac{d}{2} +1\end{matrix},y\right)_{x=0}\Bigg]\,, \nonumber
\end{align}
which, like the anomalous dimensions \eqref{scalar_anal}, we expressed as an integrated product of Gauss hypergeometric functions.\footnote{While in general the OPE coefficients \eqref{OPEomega} are given by an integrated product of Gauss hypergeometric functions, in some simple cases the integral can be evaluated. E.g. for $\tau=d-2$, as relevant for the $O\left(N\right)$ model, we have
\begin{equation}
\bar{a}^{\text{anal.}}_{0,\ell^\prime|d-2,0}+\bar{b}^{\text{anal.}}_{0,\ell^\prime|d-2,0}=\frac{\Gamma \left(\frac{d-1}{2}\right) 2^{d-2 \Delta -\ell^\prime} \Gamma (\ell^\prime+\Delta ) \Gamma \left(-\frac{d}{2}+\ell^\prime+2 \Delta \right)}{\Gamma \left(\frac{d}{2}-1\right) \Gamma \left(-\frac{d}{2}+\Delta +1\right)^2 \Gamma \left(\frac{d}{2}+\ell^\prime\right) \Gamma \left(\ell^\prime+\Delta -\frac{1}{2}\right)}\,\Xi_{\ell^\prime}.
\end{equation}
 } We also introduced the coefficient $\Xi_\ell$, which reads:
\begin{align}
    \Xi_\ell\equiv-\psi ^{(0)}\left(\tfrac{d}{2}+\ell\right)+\psi ^{(0)}(2 (\ell+\Delta ))-\psi ^{(0)}(\ell+2 \Delta -1)+\psi ^{(0)}\left(\tfrac{\tau }{2}\right)+\gamma\,.
\end{align}
To the best of our knowledge, this result for the analytic in spin double-twist OPE coefficients is new. 

For $\ell^\prime=0$ the above analytic in spin OPE data receives a correction from the contributions \eqref{lpeqlfsup} with finite support in spin. The correction to the anomalous dimension \eqref{scalar_anal} was given in \eqref{gammaEll00}. In figure \ref{fig:graph1} we plot the full anomalous dimension $\gamma_{0,\ell^\prime|\tau,0}=\gamma^{\text{anal.}}_{0,\ell^\prime|\tau,1}+\gamma^{\text{n.-a.}}_{0,\ell^\prime|\tau,0}$ as a function of $\tau$ for some fixed values of $\Delta$, $d$ and $\ell^\prime$.

The correction to the OPE coefficient \eqref{OPEomega} is instead given by:
\begin{align}
 \bar{a}^{\text{n.-a.}}_{0,0|\tau,0}+\bar{b}^{\text{n.-a.}}_{0,0|\tau,0}&=\omega_{0,0}\left[\sum^\infty_{m=0}\mathfrak{p}_{\tau,\ell^\prime=0|m}(s|2\Delta)\right]\\
 &=\frac{\left(\gamma \, d (\tau -2 \Delta )+d+4 \Delta  (\gamma  \Delta -1)-\gamma\,  \tau ^2\right) \Gamma (\tau ) \Gamma \left(2 \Delta -\frac{d}{2}\right) \Gamma \left(-\frac{d}{2}+\tau +1\right)}{2 \Gamma \left(\frac{\tau }{2}\right)^4 \Gamma \left(\Delta -\frac{\tau }{2}+1\right)^2 \Gamma \left(-\frac{d}{2}+\Delta +\frac{\tau }{2}+1\right)^2}\,.\nonumber
\end{align}

Although in the above we focussed on the OPE data of the double-twist operators with $n=0$, results for the full family of double-twist operators for general integer $n\ge 0$ can be obtained in a similar way using the techniques developed in \cite{Sleight:2018ryu}, which we briefly reviewed around equation \eqref{func}.
 
\paragraph{Vector exchange $\ell=1$.} For exchanges of odd spin, only when the external scalars carry colour indices is the fully crossing symmetric Polyakov block \eqref{co_sum} non-vanishing. The cyclic Polyakov blocks for $\ell=1$ read:
\begin{subequations}
\begin{align}
\mathcal{E}^{(1234)}_{\tau,1|m}&=\mathfrak{c}_{m}^{(1)}\left[\underbrace{\frac{2(2\Delta- m- t )-\tau }{\tau\,  (2 m+s+t+\tau -2 \Delta)}}_{\mathfrak{A}^{({\sf u})}_{\tau,1|m}(s|t)}+\underbrace{\frac{2 (m+ s)+\tau }{\tau  (2 m-t+\tau )}}_{\mathfrak{p}^{(1234)}_{\tau,1|m}(s,t)}+\underbrace{\frac{2}{\tau }}_{\mathfrak{q}^{(1234)}_{\tau,1|m}(s,t)}\right]\,,\\
\mathcal{E}^{(1423)}_{\tau,1|m}&=\mathfrak{c}_{m}^{(1)}\left[\frac{2(m+t-2\Delta)+\tau }{\tau\,  ( s+2\Delta -\tau-2m )}+\underbrace{\frac{2 (m+ s)+\tau }{ \tau  (t-\tau-2 m )}}_{\mathfrak{p}^{(1342)}_{\tau,1|m}(s,t)=-\mathfrak{p}^{(1234)}_{\tau,1|m}(s,t)}+\underbrace{\frac{4}{\tau}}_{\mathfrak{q}^{(1342)}_{\tau,1|m}(s,t)}\right]\,,\\
\mathcal{E}^{(1342)}_{\tau,1|m}&=\mathfrak{c}_{m}^{(1)}\left[\frac{2 ( m+t-2 \Delta)+\tau }{\tau  (-2 \Delta +2 m-s+\tau )}+\frac{2 (m+t-2 \Delta)+\tau }{\tau  (s+t+\tau-2 \Delta +2 m )}\right]\,,
\end{align}
\end{subequations}
where, as in \eqref{s-expansion1234}, we separated the polynomial terms in the Mellin variable $s$ from the single poles. As before, without loss of generality we focus on extracting the OPE data from $\mathcal{E}^{(1234)}_{\tau,1}$ in the following.

We first consider the analytic in spin OPE data. Such contributions to the anomalous dimensions of the double-twist operators were already extracted in \cite{Sleight:2018ryu}, where for the leading twist ($n=0$) operators we have
\begin{align}\label{spin1_anal}
  -\frac12 a^{(0)}_{0,\ell}\gamma^{\text{anal.}}_{0,\ell^\prime|\tau,1}&= {\omega}^{\pl}_{0,\ell^\prime}\left[\sum^\infty_{m=0}\mathfrak{A}^{(\sf{u})}_{\tau,\ell=1|m}(s,2\Delta)\right]\\ \nonumber
  &=-\frac{\Gamma \left(\frac{\tau +3}{2}\right) 2^{-2 \Delta -\ell^\prime+\tau +1} \Gamma (\ell^\prime+\Delta ) \Gamma (\ell^\prime+2 \Delta -1)}{(-d+\tau +2) \ell^\prime!\, \Gamma \left(\frac{\tau }{2}+1\right) \Gamma \left(-\frac{d}{2}+\tau +2\right) \Gamma \left(\Delta -\frac{\tau }{2}\right)^2 \Gamma \left(\ell^\prime+\Delta -\frac{1}{2}\right)}\\&\nonumber\times\int_0^1 dy\left[\sum_{k=0}^2\frac{a_k(\ell^\prime)}{(1-y)^k}\right](1-y)^{\ell^\prime} y^{-\frac{d}{2}+\tau +1} \, _2F_1\left(\begin{matrix}\tfrac{2-d+\tau}{2},\tfrac{\tau-2 \Delta +2}{2} \\\tau-\frac{d}{2} +2\end{matrix};y\right)^2\,,
\end{align}
with
\begin{subequations}
\begin{align}
a_0(\ell^\prime)&=-\frac{(d+2 \ell^\prime) (\Delta +\ell^\prime) (2 \Delta +\ell^\prime-1) (2 \Delta +\ell^\prime) \Gamma \left(-\frac{d}{2}+\ell^\prime+2 \Delta +1\right)}{(2 \Delta +2 \ell^\prime-1) \Gamma (\ell^\prime+2 \Delta +1)}\,,\\
a_1(\ell^\prime)&=\frac{\ell^\prime (-d+2 \Delta +2) (2 \Delta +\ell^\prime-1) \Gamma \left(-\frac{d}{2}+\ell^\prime+2 \Delta \right)}{\Gamma (\ell^\prime+2 \Delta )}\,,\\
a_2(\ell^\prime)&=\frac{\ell^\prime(\ell^\prime-1) (\Delta +\ell^\prime-1) (4 \Delta +2 \ell^\prime-2-d) \Gamma \left(-\frac{d}{2}+\ell^\prime+2 \Delta -1\right)}{(2 \Delta +2 \ell^\prime-1) \Gamma (\ell^\prime+2 \Delta -1)}\,.
\end{align}
\end{subequations}
As for the scalar exchange we employed the representation of the Wilson function given by an integrated product of Gauss hypergeometric functions. It's important to note however that there is a technical subtlety in using this representation for $\ell^\prime=1$ and $\ell^\prime=0$ which is that, while this representation is manifestly analytic in spin, setting $\ell^\prime=1$ and $\ell^\prime=0$ does not commute with the integral over $y$ owing to the zeros of the functions $a_{1}\left(\ell^\prime\right)$ and $a_{2}\left(\ell^\prime\right)$ at these values which are compensated by the divergence of the integrand as $y \sim 1$. This can be resolved by integrating by parts, in particular using the identity
\begin{multline}\label{byparts}
    \int_{0}^1 dy\, a_k(\ell^\prime)\,(\ell^\prime-k+1)_k\,(1-y)^{\ell^\prime-k} f(y)\\=\int_{0}^1 dy\, a_k(\ell^\prime)(1-y)^{\ell^\prime}\pl_{y}^k\,\left[f(y)\right]=\ell^\prime!\,a_k(\ell^\prime)\,\pl_y^{k-\ell^\prime-1}f(y)\Big|_{y=0}\,,
\end{multline}
which is valid for $k,\ell^\prime\in\mathbb{N}$ and $k>\ell^\prime$ and where we have assumed that $f^{(q)}(0)=0$ which can always be ensured by choosing $\tau$ large enough. 

For $\ell^\prime=1$ we obtain
\begin{align}
  -\frac12 a^{(0)}_{0,1}\gamma^{\text{anal.}}_{0,1|\tau,1}&= {\omega}^{\pl}_{0,1}\left[\sum^\infty_{m=0}\mathfrak{A}^{(\sf{u})}_{\tau,\ell=1|m}(s,2\Delta)\right]\\ \nonumber
  &=-\frac{2^{\tau -2 \Delta } \Gamma (\Delta ) \Gamma (\Delta +1)^2 \Gamma \left(\frac{\tau +3}{2}\right) \Gamma \left(-\frac{d}{2}+2 \Delta +1\right) \Gamma \left(-\frac{d}{2}+\tau +2\right)}{(-d+\tau +2) \Gamma \left(\Delta +\frac{3}{2}\right) \Gamma \left(\frac{\tau }{2}+1\right)^3 \Gamma \left(\Delta -\frac{\tau }{2}\right)^2 \Gamma \left(-\frac{d}{2}+\Delta +\frac{\tau }{2}+1\right)^2}\nonumber\\
&-\frac{\Gamma \left(\frac{\tau +3}{2}\right) 2^{\tau-2 \Delta} \Gamma (\Delta+1 ) \Gamma (2 \Delta)}{(-d+\tau +2) \Gamma \left(\frac{\tau }{2}+1\right) \Gamma \left(-\frac{d}{2}+\tau +2\right) \Gamma \left(\Delta -\frac{\tau }{2}\right)^2 \Gamma \left(\Delta +\frac{1}{2}\right)}\\&\nonumber\times\int_0^1 dy\left[a_0(1)+\frac{a_1(1)}{1-y}\right](1-y) y^{-\frac{d}{2}+\tau +1} \, _2F_1\left(\begin{matrix}\tfrac{2-d+\tau}{2},\tfrac{\tau-2 \Delta +2}{2} \\\tau-\frac{d}{2} +2\end{matrix};y\right)^2\,.
\end{align}
and for $\ell^\prime=0$:
\begin{align}
  -\frac12 a^{(0)}_{0,0}\gamma^{\text{anal.}}_{0,0|\tau,1}&= {\omega}^{\pl}_{0,0}\left[\sum^\infty_{m=0}\mathfrak{A}^{(\sf{u})}_{\tau,\ell=1|m}(s,2\Delta)\right]\\ \nonumber
  &=\frac{2^{\tau -2 \Delta } \Gamma (\Delta )^3 \Gamma \left(\frac{\tau +3}{2}\right) \left(d (\Delta +\tau )-4 \Delta ^2-\tau  (\tau +2)\right) \Gamma \left(2 \Delta -\frac{d}{2}\right) \Gamma \left(-\frac{d}{2}+\tau +2\right)}{(-d+\tau +2) \Gamma \left(\Delta +\frac{1}{2}\right) \Gamma \left(\frac{\tau }{2}+1\right)^3 \Gamma \left(\Delta -\frac{\tau }{2}\right)^2 \Gamma \left(-\frac{d}{2}+\Delta +\frac{\tau }{2}+1\right)^2}\nonumber\\
&-\frac{\Gamma \left(\frac{\tau +3}{2}\right) 2^{\tau-2 \Delta} \Gamma (\Delta+1 ) \Gamma (2 \Delta)}{(-d+\tau +2) \Gamma \left(\frac{\tau }{2}+1\right) \Gamma \left(-\frac{d}{2}+\tau +2\right) \Gamma \left(\Delta -\frac{\tau }{2}\right)^2 \Gamma \left(\Delta +\frac{1}{2}\right)}\\&\nonumber\times\int_0^1 dy\,a_0(0)\,y^{-\frac{d}{2}+\tau +1} \, _2F_1\left(\begin{matrix}\tfrac{2-d+\tau}{2},\tfrac{\tau-2 \Delta +2}{2} \\\tau-\frac{d}{2} +2\end{matrix};y\right)^2\,.
\end{align}

On top of the above analytic in spin contributions to the anomalous dimensions, we must also include the non-analytic correction for $\ell^\prime=1$ and $\ell^\prime=0$. For $\ell^\prime=1$ this was given in \eqref{ContactGen}. For $\ell^\prime=0$ it is: 
\begin{align}
-\frac12 a^{(0)}_{0,0}\gamma^{\text{n.-a.}}_{0,0|\tau,1}&=  \omega^{\pl}_{0,\ell=0}\left[\sum_m\mathfrak{p}^{(1234)}_{\tau,1|m}(s|2\Delta)+\mathfrak{q}^{(1234)}_{\tau,1|m}(s|2\Delta)\right]\\ \nonumber
&=-\frac{3 (2 \Delta +1) (2 \Delta -\tau ) (2-d+2 \Delta +\tau)}{4 \Delta  (d-4 \Delta )}\,a_{0,1}^{(0)}\,\gamma^{\text{n.-a.}}_{0,1|\tau,1},\
\end{align}
where for concision we expressed the result in terms of the correction for $\ell^\prime=1$, given by \eqref{ContactGen}.

\paragraph{Spin-2 exchange $\ell=2$.} In this case the analytic in spin contribution to the anomalous dimension of the leading ($n=0$) double-twist operators reads \cite{Sleight:2018ryu}:
\begin{align}\label{spin2anal}
 -\frac12 a^{(0)}_{0,\ell^\prime}\gamma^{\text{anal.}}_{0,\ell^\prime|\tau,2}&=\omega^{\pl}_{n,\ell^\prime}\left[\sum^\infty_{m=0}\mathfrak{A}^{({\sf u})}_{\tau,\ell=2|m}(s|2\Delta)\right]\\ \nonumber
 &=\frac{\Gamma (\tau +4) \Gamma (d-\tau -3) \Gamma (\ell^\prime+\Delta )^2 \Gamma \left(-\frac{d}{2}+\ell^\prime+2 \Delta \right)}{\ell^\prime! \Gamma \left(\frac{\tau }{2}+2\right)^2 \Gamma (d-\tau -1) \Gamma \left(-\frac{d}{2}+\tau +3\right) \Gamma \left(\Delta -\frac{\tau }{2}\right)^2 \Gamma (2 (\ell^\prime+\Delta ))}\\&\nonumber \hspace*{2cm}\times\int_0^1 dy\left[\sum_{k=0}^4\frac{a_k(\ell^\prime)}{(1-y)^k}\right](1-y)^{\ell^\prime} y^{-\frac{d}{2}+\tau +2} \, _2F_1\left(\begin{matrix}\tfrac{2-d+\tau}{2},\tfrac{\tau-2 \Delta +2}{2} \\\tau-\frac{d}{2} +3\end{matrix};y\right)^2\,,
\end{align}
with
\begin{subequations}
\begin{align}
    a_0(\ell^\prime)&=\frac{2^{\ell^\prime-6}(d-1) (d+2 \ell^\prime) (d+2 \ell^\prime+2) (\Delta +\ell^\prime) (\Delta +\ell^\prime+1) (d-2 (2 \Delta +\ell^\prime)) (d-2 (2 \Delta +\ell^\prime+1))}{d (2 \Delta +2 \ell^\prime+1)}\,,\\
    a_1(\ell^\prime)&=-\frac{2^{\ell^\prime-4} \ell^\prime (d+2 \ell^\prime) (\Delta +\ell^\prime) (-d+4 \Delta +2 \ell^\prime) (d (d-2 \Delta -3)+2 (\Delta +\ell^\prime))}{d}\,,\\
    a_2(\ell^\prime)&=-\frac{2^{\ell^\prime-3}\ell^\prime(\ell^\prime-1) (2 \Delta +2 \ell^\prime-1)}{d (2 \Delta +2\ell^\prime-3) (2 \Delta +2 \ell^\prime+1)}\Big(d^3 \left(3 \Delta -3 \left(\Delta ^2+2 \Delta  \ell^\prime+(\ell^\prime-1) \ell^\prime\right)+2\right)\\\nonumber&+d^2 \left(12 \Delta ^3-19 \Delta +\Delta ^2 (24 \ell^\prime-1)+2 \Delta  \ell^\prime (6 \ell^\prime+5)+11 (\ell^\prime-1) \ell^\prime-8\right)\\\nonumber
    &+2 d \Big(\Delta  (15-4 \Delta  (\Delta  (\Delta +2)-2))+2 \left(\ell^\prime\right)^4+(8 \Delta -4) \left(\ell^\prime\right)^3+\left(6 \Delta ^2-22 \Delta -6\right) \left(\ell^\prime\right)^2\\\nonumber
    &-2 (2 \Delta -1) (\Delta  (\Delta +7)+4) \ell^\prime+5\Big)+4 \Big((\Delta -1)^3 (2 \Delta +1)+3 \left(\ell^\prime\right)^4+6 (2 \Delta -1) \left(\ell^\prime\right)^3\\\nonumber
    &+(\Delta  (17 \Delta -18)+2) \left(\ell^\prime\right)^2+\Delta  (\Delta  (10 \Delta -17)+4) \ell^\prime+\ell^\prime\Big)\Big)\,,\\
    a_3(\ell^\prime)&=-\frac{2^{\ell^\prime-2}\ell^\prime (\ell^\prime-1) (\ell^\prime-2) (\Delta +\ell^\prime-1) (d (-d+2 \Delta +3)+2 (\Delta +\ell^\prime-1))}{d}\,,\\
    a_4(\ell^\prime)&=\frac{2^{\ell^\prime-2}(d-1)  \ell^\prime(\ell^\prime-1) (\ell^\prime-2) (\ell^\prime-3) (\Delta +\ell^\prime-2) (\Delta +\ell^\prime-1)}{d (2 \Delta +2 \ell^\prime-3)}\,.
\end{align}
\end{subequations}
As for the vector exchange considered in the previous example, while this expression is manifestly analytic in spin, extra care needs to be taken for $\ell^\prime=0,1,2,3$ where the functions $a_{1}\left(\ell^\prime\right)$, ..., $a_{4}\left(\ell^\prime\right)$ have zeros which are compensated the divergence of the integrand as $y \sim 1$. As before, this issue can be removed by integrating by parts \eqref{byparts}. For $\ell^\prime=0,1,2,3$ we need to add the following boundary terms:
\begin{itemize}
    \item $\ell^\prime=3$:
    \begin{align}
    \delta_{\ell^\prime,3}\,\frac{(d-1) (\Delta +1) (\Delta +2) 2^{-2 \Delta +\tau -1} \Gamma (\Delta +1)^2 \Gamma (\Delta +3) \Gamma \left(\frac{\tau +5}{2}\right) \Gamma \left(-\frac{d}{2}+2 \Delta +3\right) \Gamma \left(-\frac{d}{2}+\tau +3\right)}{d (2 \Delta +3) (d-\tau -3) (d-\tau -2) \Gamma \left(\Delta +\frac{7}{2}\right) \Gamma \left(\frac{\tau }{2}+2\right)^3 \Gamma \left(\Delta -\frac{\tau }{2}\right)^2 \Gamma \left(-\frac{d}{2}+\Delta +\frac{\tau }{2}+2\right)^2}\,.
    \end{align}
    \item $\ell^\prime=2$:
    \begin{align}
    \delta_{\ell^\prime,2}&\,\left(-d^2 (3 \Delta +\tau +4)+d (\Delta  (8 \Delta +15)+\tau  (\tau +5)+12)+4 \Delta  (2 \Delta +3)-\tau  (\tau +4)\right)\nonumber\\
    &\,\frac{2^{-2 \Delta +\tau -1} \Gamma (\Delta +1) \Gamma (\Delta +2)^2 \Gamma \left(\frac{\tau +5}{2}\right) \Gamma \left(-\frac{d}{2}+2 \Delta +2\right) \Gamma \left(-\frac{d}{2}+\tau +3\right)}{d (1+2 \Delta) (-d+\tau +2) (-d+\tau +3) \Gamma \left(\Delta +\frac{5}{2}\right) \Gamma \left(\frac{\tau }{2}+2\right)^3 \Gamma \left(\Delta -\frac{\tau }{2}\right)^2 \Gamma \left(\tfrac{-d+2 \Delta +\tau +4}{2}\right)^2}\,,
    \end{align}
    \item $\ell^\prime=1$:
    \begin{align}
    \delta_{\ell^\prime,1}&\,\frac{2^{-2 \Delta +\tau -4} \Gamma (\Delta -2) \Gamma (\Delta +1)^2 \Gamma \left(\frac{\tau +5}{2}\right) \Gamma \left(-\frac{d}{2}+2 \Delta +1\right) \Gamma \left(-\frac{d}{2}+\tau +3\right)}{\Gamma \left(\Delta +\frac{3}{2}\right) \Gamma \left(\frac{\tau }{2}+2\right)^3 \Gamma \left(\Delta -\frac{\tau }{2}\right)^2 \Gamma \left(-\frac{d}{2}+\Delta +\frac{\tau }{2}+2\right)^2}\\\nonumber
    &\times\frac{4 (\Delta -2) (\Delta -1)}{d (2 \Delta +3) (d-\tau -3) (d-\tau -2)}\,\Big(2 \Big(d^3 (\Delta +1) (\Delta  (3 \Delta +8)+6)\\\nonumber
    &-d^2 (\Delta  (\Delta  (\Delta  (16 \Delta +65)+93)+66)+30)+2 d (\Delta  (\Delta  (\Delta  (4 \Delta  (2 \Delta +9)+67)+61)+38)+24)\\\nonumber
    &-8 (\Delta +1) \left(2 \Delta ^4+5 \Delta ^3-3 \Delta +3\right)\Big)+(2 \Delta +3) \tau ^2 \Big(8 (d+1) \Delta ^2-2 (d-3) d \Delta +(d-9) (d-4) d\\\nonumber
    &+4 (\Delta -6)\Big)+2 (d-4) (2 \Delta +3) \tau  \Big(-4 (d+1) \Delta ^2+(d-3) d \Delta +2 (d-3) d-2 \Delta +4\Big)\\\nonumber
    &+(d-1) (2 \Delta +3) \tau ^4-2 (d-4) (d-1) (2 \Delta +3) \tau ^3\Big)\,
    \end{align}
    \item $\ell^\prime=0$:
    {\allowdisplaybreaks
    \begin{align}
    \delta_{\ell^\prime,0}&\,\frac{2^{-2 \Delta +\tau -3} \Gamma (\Delta )^3 \Gamma \left(\frac{\tau +5}{2}\right) \Gamma \left(2 \Delta -\frac{d}{2}\right) \Gamma \left(-\frac{d}{2}+\tau +3\right)}{d (-d+\tau +2) (-d+\tau +3) \Gamma \left(\Delta +\frac{3}{2}\right) \Gamma \left(\frac{\tau }{2}+2\right)^3 \Gamma \left(\Delta -\frac{\tau }{2}\right)^2 \Gamma \left(-\frac{d}{2}+\Delta +\frac{\tau }{2}+2\right)^2}\\\nonumber
    &\times\,d^4 \left(-\left(2 \Delta ^4+2 \Delta ^3 (\tau +3)+2 \Delta ^2 (\tau +2)^2+\Delta  (\tau +2)^2 (2 \tau +1)+\tau  (\tau +2)^2\right)\right)\\\nonumber
    &+d^3 \Big(16 \Delta ^5+2 \Delta ^4 (8 \tau +27)+6 \Delta ^3 (\tau  (3 \tau +13)+15)+2 \Delta ^2 (\tau +2)^2 (2 \tau +1)\\\nonumber
    &+\Delta  (\tau +2)^2 \left(6 \tau ^2+20 \tau +5\right)+(\tau +2)^2 (3 \tau  (\tau +3)+4)\Big)\\\nonumber
    &+d^2 \Big(-32 \Delta ^6-32 \Delta ^5 (\tau +4)-4 \Delta ^4 (4 \tau  (\tau +8)+57)-2 \Delta ^3 (\tau  (\tau  (16 \tau +91)+174)+114)\\\nonumber
    &-2 \Delta ^2 (\tau +2) \left(\tau ^3-16 \tau -28\right)-\Delta  (\tau +2)^3 (3 \tau  (2 \tau +9)+8)-(\tau +1) (\tau +2)^3 (3 \tau +10)\Big)\\\nonumber
    &+d \Big(32 \Delta ^6+16 \Delta ^5 (2 \tau  (\tau +5)+15)+16 \Delta ^4 (\tau  (4 \tau +17)+19)+4 \Delta ^3 (\tau +2)^2 (4 \tau  (\tau +2)+1)\\\nonumber
    &-2 \Delta ^2 (\tau +2)^2 (\tau +4) (3 \tau +8)+\Delta  (\tau +2)^4 (2 \tau  (\tau +7)+9)+(\tau +2)^4 (\tau  (\tau +7)+8)\Big)\\\nonumber
    &+(-2 \Delta -1) (\tau +2)^2 \left((\tau +2)^2-4 \Delta ^2\right)^2\,.
    \end{align}}
\end{itemize}

For $\ell^\prime \leq 2$ the above analytic in spin contributions to the double-twist anomalous dimensions receive a non-analytic correction. For $\ell^\prime=2$ this is given by \eqref{gammaEll}. For $\ell^\prime=1$ we have 
\begin{align}
 -\frac12 a^{(0)}_{0,1}\gamma^{\text{n.-a.}}_{0,1|\tau,2}&= \omega^{\pl}_{0,\ell=1}\left[\sum_m\mathfrak{p}^{(1234)}_{\tau,2|m}(s|t)+\mathfrak{q}^{(1234)}_{\tau,2|m}(s|t)\right]\\
 &=-\frac{(2 \Delta +3) (2 \Delta -\tau ) (d-2 \Delta -\tau -4)}{2(\Delta +1) (d-4 \Delta -2)}\,a^{(0)}_{0,1}\gamma^{\text{n.-a.}}_{0,1|\tau,1}\,,
\end{align}
where as before for concision we expressed the results in terms of \eqref{gammaEll}. For $\ell^\prime=0$ we have:
\begin{align}
   -\frac12 a^{(0)}_{0,0}\gamma^{\text{n.-a.}}_{0,0|\tau,2}&=\omega^{\pl}_{0,\ell=0}\left[\sum_m\mathfrak{p}^{(1234)}_{\tau,2|m}(s|t)+\mathfrak{q}^{(1234)}_{\tau,2|m}(s|t)\right]\\
   &=-\frac{(2 \Delta +3)}{16 (\Delta +1)} \Bigg[-\tfrac{8 (2 \Delta +1) (2 \Delta -\tau -2) (2 \Delta -\tau )}{d-4 \Delta -2}\\\nonumber
    &+\tfrac{(2 \Delta +1) \left(d^2 (\tau +1)-d (\tau +1) (\tau +4)+(\tau +2)^2\right) (2 \Delta -\tau -4) (2 \Delta -\tau ) (-2 \Delta +\tau +2)^2}{d \Delta  (\tau +1) (d-4 \Delta -2) (d-4 \Delta ) (d-\tau -3)}\nonumber\\&\nonumber
    +\tfrac{2 (2 \Delta +1) (2 \Delta -\tau -2) (2 \Delta -\tau ) \left(d (\tau +1) (8 \Delta -\tau -2)+(\tau +2)^2\right)}{d \Delta  (\tau +1) (d-4 \Delta -2)}\\\nonumber
    &+\tfrac{(2 \Delta +1) \left(32 \Delta ^2 (\tau +1)-16 \Delta  \tau  (\tau +1)+\tau ^2 (\tau +2)\right)}{\Delta  (\tau +1)}-8 (2 \Delta +1) (3 \Delta -\tau +1)+8 (\Delta +1)^2\Bigg]\,a^{(0)}_{0,0}\,\gamma^{\text{n.-a.}}_{0,0|\tau,0}\,.
\end{align}
As a check of the above expressions, we have compared the full crossing symmetric solution obtained by summing the cyclic amplitudes with the result obtained in \cite{Alday:2017gde} for $\tau=2$ and $d=4$, finding agreement up to a choice of contact $\phi^4$ term which was not fixed in \cite{Alday:2017gde}.

\subsection{Relation with dual bootstrap functionals}
\label{subsec::abfunc}

Owing to the manifest crossing symmetry of cyclic exchange amplitude, the direct channel decomposition discussed so far turns out to be intimately related to the concept of dual functionals defined on the space of bootstrap vectors (see \cite{Poland:2018epd} for a review):
\begin{align}
    F_{\tau,\ell}(u,v)=g_{\tau,\ell}(u,v)-g_{\tau,\ell}(v,u)\,,\label{bootvec}
\end{align}
where in particular one is considering the crossing relation associated to the exchange of legs $2$ and $4$ which in our conventions maps the $\sf{s}$ and $\sf{u}$-channels into each other.

In \cite{Rychkov:2017tpc} it was clarified that a well-defined bootstrap functional has to satisfy key finiteness and swapping condition to give rise to well-defined sum rules. In particular, given a linear functional $W$ on the space of bootstrap vectors \eqref{bootvec}, apart from the condition that its action on conformal blocks and four-point correlator should be finite, one should also require the following key convergence condition:
\begin{align}
    W\left[\sum_{\mathcal{O}}a_{\phi\phi\mathcal{O}}\,g_{\tau_{\mathcal{O}},\ell_{\mathcal{O}}}\right]=\sum_{\mathcal{O}}a_{\phi\phi\mathcal{O}}W\left[g_{\tau_{\mathcal{O}},\ell_{\mathcal{O}}}\right]\,,
\end{align}
with the second sum being absolutely convergent. Leaving the above conditions aside for a moment it is standard to define the following basis of the dual space to mean-field theory bootstrap vectors via:
\begin{align}\label{dual}
    W^{(1)}_{m,\ell}[F_{\tau_1+\tau_2+2n,\ell^\prime}]&=\delta_{n,m}\delta_{\ell,\ell^\prime}\,,&
    W^{(2)}_{m,\ell}[F_{\tau_3+\tau_4+2n,\ell^\prime}]&=\delta_{n,m}\delta_{\ell,\ell^\prime}\,.
\end{align}
If such dual functionals satisfying the finiteness and swapping conditions exist, the crossing symmetry of the cyclic Polyakov block ${\cal E}^{(1234)}_{\tau,\ell^\prime}$ with respect to the ${
\sf s}$- and ${\sf u}$-channels implies the following sum rules:
\begin{multline}
    W^{(i)}_{m,\ell}[F_{\tau,\ell'}(u,v)]+\sum_{n,J=0}^\infty \omega^{(1)}_{n,J}\left[{\cal E}^{(1234)}_{\tau,\ell^\prime}\right]\, W^{(i)}_{m,\ell}[F_{\tau_1+\tau_2+2n,J}(u,v)]\\+\sum_{n,\ell^\prime=0}^\infty \omega^{(2)}_{n,J}\left[{\cal E}^{(1234)}_{\tau,\ell^\prime}\right]\, W^{(i)}_{m,\ell}[F_{\tau_3+\tau_4+2n,J}(u,v)]=0\,,
\end{multline}
which then gives the equality:
\begin{align}
    W^{(i)}_{n,\ell}[F_{\tau,\ell^\prime}]+\omega^{(i)}_{n,\ell^\prime}\left[{\cal E}^{(1234)}_{\tau,\ell^\prime}\right]=0\,.
\end{align}
This identity states the equivalence between the coefficients in the conformal block decomposition of cyclic Polyakov blocks and the dual basis of functionals to the crossing vectors $\mathcal{F}_{\tau_{min}+2n,\ell}$  .

Cyclic Polyakov blocks should therefore be considered as compact generating functions for dual functionals to the bootstrap vectors. Our prescription to uniquely fix cyclic Polyakov blocks therefore automatically translates into a prescription to define dual functionals to bootstrap vectors in general $d$. We leave some applications of our result which include revisiting the bootstrap of the Wilson-Fisher fixed point in the $\epsilon$-expansion to higher orders in $\epsilon$ for the future.

\section*{Acknowledgments}
C.S. gratefully acknowledges Universit\'e Libre de Bruxelles, Princeton University and the  University of  Naples  Federico  II for support and hospitality during various stages of this work. The research of C.S. is supported by the European Union's Horizon 2020 research and innovation programme under the Marie Sk\l odowska-Curie grant agreement No 793661 and, until October 2018, by a Marina Solvay Fellowship. M.T. thanks Caltech and the Simons Bootstrap Collaboration for hospitality and support during the Bootstrap workshop in July 2018, where the initial stages of this work were carried out. The research of M.T. was partially supported by the program  “Rita  Levi  Montalcini”  of the MIUR (Minister for Instruction, University and Research), the INFN initiative STEFI and the European Union's Horizon 2020 research and innovation programme under the Marie Sklodowska-Curie grant agreement No 747228.

\begin{appendix}

\section{Mack polynomials}\label{Mack}

The kinematic polynomials $\mathcal{Q}_{\tau,\ell|m}$ are entirely defined by the Mellin representation of the corresponding Conformal Partial Wave (see e.g. \cite{Sleight:2018epi}), 
\begin{subequations}
\begin{align}\label{kineQ}
\mathcal{Q}_{\tau,\ell|m}(s)&=\text{Res}_{t=\tau+2m}\left({}^{(\sf{s})}\mathcal{F}_{\tau,\ell}(s,t)\right)
\\ \label{mrepexsc}
    {}^{(\sf s)}\mathcal{F}_{\tau,\ell}\left(s,t\right)&= \mathcal{C}_{\tau,\ell}(\tau_i)\,\Omega_\ell(t)\, {}^{(\sf s)}P_{\tau,\ell}(s,t)\,,
\end{align}
\end{subequations}
where we have defined the following factors:
\begin{subequations}
\begin{align}\label{cltau}
    \mathcal{C}_{\tau,\ell}(\tau_i)&=\frac{4^{-\ell} (\ell+\tau -1)_\ell \Gamma (2 
    \ell+\tau )}{\Gamma \left(\tfrac{d-2 (\ell+\tau )}{2}\right) \Gamma \left(\tfrac{2 \ell+\tau +\tau_1-\tau_2}{2}\right) \Gamma \left(\tfrac{2 \ell+\tau -\tau_1+\tau_2}{2}\right) \Gamma \left(\tfrac{2 \ell+\tau +\tau_3-\tau_4}{2}\right) \Gamma \left(\tfrac{2 \ell+\tau -\tau_3+\tau_4}{2}\right)},
    \\ \label{omegalt}
    \Omega_\ell(t)&=\frac{\Gamma \left(\tfrac{\tau -t}{2}\right) \Gamma \left(\tfrac{d-2 \ell-t-\tau}{2}\right)}{\Gamma \left(\tfrac{-t+\tau_1+\tau_2}{2}\right) \Gamma \left(\tfrac{-t+\tau_3+\tau_4}{2}\right)
    }\,.
\end{align}
\end{subequations}
and $P_{\tau,\ell}(s,t)$ is known in the literature as a Mack polynomial \cite{Mack:2009mi}, which can be expressed in the following form \cite{Sleight:2018epi}:
\begin{align}\label{MackPscalar}
    P^{(\sf s)}_{\tau,\ell^\prime}(s,t|\tau_1,\tau_2,\tau_3,\tau_4)=
    \sum_{k=0}^{[\ell^\prime/2]}c_{\ell^\prime,k}\left[\sum_{\sum_ir_i=\ell^\prime-2k}p^{(r_1,r_2,r_3,r_4)}_{k,\ell^\prime}(s,t|\tau_1,\tau_2,\tau_3,\tau_4)\right]\,,
\end{align}
where $P_{\tau,\ell^\prime}(s,t)\sim s^{\ell^\prime}+\ldots$, the coefficients $c_{\ell^\prime,k}$ are the Gegenbauer expansion coefficients
\begin{equation}
    c_{\ell,k}=\frac{(-4)^{-k} \ell! \Gamma \left(\frac{d}{2}-k+\ell-1\right)}{k! (\ell-2 k)! \Gamma \left(\frac{d}{2}+\ell-1\right)}\,,
\end{equation}
and we introduced
\begin{multline}
    p^{(r_i)}_{k,\ell^\prime}(s,t|\tau_1,\tau_2,\tau_3,\tau_4)=\left(\frac{\tau -t}{2}\right)_k \left(\frac{d-2 \ell^\prime-t-\tau}{2}\right)_k\\\times\,\frac{\left(\frac{-d+\tau +\tau_3-\tau_4+2}{2} \right)_{k+r_1+r_2} \left(\frac{-d+\tau -\tau_3+\tau_4+2}{2}\right)_{k+r_3+r_4}}{\left(\frac{\tau -\tau_1+\tau_2}{2}\right)_{k+r_1+r_3} \left(\frac{\tau +\tau_1-\tau_2}{2}\right)_{k+r_2+r_4}}\,\bar{p}^{(r_i)}(s,t|\tau_1,\tau_2,\tau_3,\tau_4),
\end{multline}
with
\begin{multline}
    \bar{p}^{(r_i)}(s,t|\tau_1,\tau_2,\tau_3,\tau_4)=\frac{(-1)^{r_1+r_4}}{2^{r_1+r_2+r_3+r_4}}(r_1,r_2,r_3,r_4)!\\\times\, \left(\frac{s+t}{2}\right)_{r_1} \left(\frac{-s+\tau_3-\tau_4}{2}\right)_{r_2} \left(\frac{-s-\tau_1+\tau_2}{2}\right)_{r_3} \left(\frac{s+t+\tau_1-\tau_2-\tau_3+\tau_4}{2}\right)_{r_4}\,.
\end{multline}
We often find it convenient to factor out the following normalisation coefficient from \eqref{kineQ}:
\begin{equation}\label{clm}
    \mathfrak{c}_{m}^{(\ell)}=\frac{(-1)^m \Gamma (2 \ell+\tau ) \Gamma \left(\tfrac{d-2 (\ell+m+\tau )}{2}\right)}{m!\, \Gamma \left(\frac{\tau }{2}\right)^2 \Gamma \left(\ell+\frac{\tau }{2}\right)^2 \Gamma \left(\frac{d}{2}-\ell-\tau \right) \Gamma \left(\Delta -\frac{\tau }{2}-m\right)^2}\,,
\end{equation}
evaluated for external operators with equal scaling dimension $\Delta$.

\section{Continuous Hahn polynomials}\label{Hahn}

In this appendix we review various properties of Continuous Hahn polynomials relevant for this work. 

Continuous Hahn polynomials \cite{andrews_askey_roy_1999} $Q_\ell^{(a,b,c,d)}(s)$ are orthogonal with respect to the Mellin-Barnes bilinear product:
\begin{equation}
    \left\langle f(s)g(s)\right\rangle_{a,b,c,d}=\int_{-i\infty}^{i\infty}\frac{ds}{4\pi i}\,\Gamma(\tfrac{s+a}2)\Gamma(\tfrac{s+b}2)\Gamma(\tfrac{c-s}2)\Gamma(\tfrac{d-s}2)\,f(s)\,g(s)\,,
\end{equation}
with normalisation
\begin{multline}\label{Qnorm}
    \left\langle Q_{\ell}^{(a,b,c,d)}(s)Q_{n}^{(a,b,c,d)}(s)\right\rangle=\delta_{\ell,n}\\\times\, \underbrace{\frac{(-1)^n 4^n n! \Gamma \left(\frac{a+c}{2}+n\right) \Gamma \left(\frac{a+d}{2}+n\right) \Gamma \left(\frac{b+c}{2}+n\right) \Gamma \left(\frac{b+d}{2}+n\right)}{\left(\tfrac{a+b+c+d}{2}+n-1\right)_n \Gamma \left(\frac{a+b+c+d}{2}+2 n\right)}}_{\mathfrak{N}^{(a,b,c,d)}_n}\,.
\end{multline}
They can be expressed explicitly in terms of a hypergeometric function ${}_3F_2$ in two equivalent forms:
\begin{subequations}
\begin{align}
    Q_{\ell}^{(a,b,c,d)}(s)&=\frac{(-2)^\ell \left(\frac{a+c}{2}\right)_\ell \left(\frac{a+d}{2}\right)_\ell}{\left(\frac{a+b+c+d}{2}+\ell-1\right)_\ell}\, _3F_2\left(\begin{matrix}-\ell,\frac{a+b+c+d}{2}+\ell-1,\frac{a+s}{2}\\\frac{a+c}{2},\frac{a+d}{2}\end{matrix};1\right)\,,\\
    Q_{\ell}^{(a,b,c,d)}(s)&=\frac{2^\ell \left(\frac{a+d}{2}\right)_\ell \left(\frac{b+d}{2}\right)_\ell}{\left(\frac{a+b+c+d}{2}+\ell-1\right)_\ell}\, _3F_2\left(\begin{matrix}-\ell,\frac{a+b+c+d}{2}+\ell-1,\frac{d-s}{2}\\\frac{a+d}{2},\frac{b+d}{2}\end{matrix};1\right)\,,
\end{align}
\end{subequations}
with unit normalisation for $s^\ell$ monomial $Q_{\ell}^{(a,b,c,d)}(s)\sim s^{\ell}+\ldots$\,. The two representations above admit the following series expansion in terms of $\left(\tfrac{-s+d}2\right)_n$ and $\left(\tfrac{s+a}2\right)_n$:
\begin{subequations}
\begin{align}
     Q_{\ell}^{(a,b,c,d)}(s)&=\sum_n\,\frac{2^{\ell} (-1)^n \binom{\ell}{n} \left(\frac{a+d}{2}+n\right)_{\ell-n} \left(\frac{b+d}{2}+n\right)_{\ell-n}}{\left(\tfrac{a+b+c+d}{2}+\ell+n-1\right)_{\ell-n}}\,\left(\tfrac{-s+d}2\right)_n\,,\label{Qtchannel}\\
     Q_{\ell}^{(a,b,c,d)}(s)&=\sum_n\,\frac{(-2)^\ell (-1)^n \binom{\ell}{n} \left(\frac{a+c}{2}+n\right)_{\ell-n} \left(\frac{a+d}{2}+n\right)_{\ell-n}}{\left(\tfrac{a+b+c+d}{2}+\ell+n-1\right)_{\ell-n}}\,\left(\tfrac{s+a}2\right)_n\,.\label{Quchannel}
\end{align}
\end{subequations}

\section{Contact term ambiguity}\label{ContactAmb}
\paragraph{Identical external legs.}

As we have seen, the final form for the exchange amplitude is ambiguous since one can always add, maintaining the same Regge behaviour of the full amplitude, a crossing symmetric solution which is given by a polynomial in the Mellin variables. The most general such crossing symmetric solution in the case of correlators with identical external legs takes the following form:
\begin{subequations}
\begin{align}\label{pl}
p_{\ell-1}(s,t)&=f_{\ell-1}(X,Y)\,,\\
X&=S\,T+T\,U+U\,S=t (2 \Delta +s)+(2 \Delta +s) (2 \Delta -s-t)+t (2 \Delta -s-t)\,,\\
Y&=S\,T\,U=t (2 \Delta +s) (2 \Delta -s-t)\,.
\end{align}
\end{subequations}
where $f_{n}(X,Y)$ is a polynomial of degree $n$ in the variables $s$ and $t$. For convenience we have also explicitly written down the manifestly crossing symmetric building blocks $X$ and $Y$ expressed in terms of the Mellin variables
\begin{align}
    S&=t\,,& T&=s+2\Delta\,,& U&=-s-t+2\Delta\,.
\end{align}
This implies that the number of independent coefficient parameterising the contact term ambiguity is:
\begin{align}
   \#= \frac{1}{72} \left(6 \ell (\ell+2)+9 (-1)^\ell-8 \sqrt{3} \sin \left(\tfrac{2 \pi  \ell}{3}\right)-8 \cos \left(\tfrac{2 \pi  \ell}{3}\right)-1\right)\,,
\end{align}
where $\sqrt{3} \sin \left(\tfrac{2 \pi  \ell}{3}\right)=0,\tfrac32,-\tfrac32$ and $ \cos \left(\tfrac{2 \pi  \ell}{3}\right)=1,-\tfrac{1}2,-\tfrac12$ with $\ell$ mod 3.

\paragraph{Non-identical external legs.}

In the case of non-identical external legs the contact term ambiguity is defined directly at the level of color ordered (cyclic) amplitudes \eqref{cyclic_exch}. In this case, the polynomial ambiguity ${p}_\ell(s,t)$ takes the following general form:
\begin{subequations}
\begin{align}
    {p}_{\ell-1}(S,U)&=f_{\ell-1}(X,Y)\,,\\
    X&=S+U\,,\\
    Y&=S\, U\,,
\end{align}
\end{subequations}
where $f_{n}(X,Y)$ is a polynomial of degree $n$ in the Mellin variables. The number of coefficients parameterising the contact term ambiguity is
\begin{equation}
    \#=\frac{1}{8} \left(2 \ell (\ell+2)+(-1)^{\ell+1}+1\right)\,.
\end{equation}

\section{Leading non-analytic piece}\label{pterm}

In this appendix we give further details on the extraction of the universal non-analytic in spin contribution \eqref{gammaEll} to the anomalous dimension of double-twist operators. This contribution comes from the $\sf{s}$-channel only and is generated by the double-trace poles at $t=2\Delta+2n$ in:
\begin{equation}
   M(s,t)= \rho(s,t)\sum_{m=0}^\infty\left(\frac{\mathcal{Q}_{\tau,\ell|m}(s)}{-t+\tau+2m}\right)\,.
\end{equation}
For the leading double-twist operators of twist $2\Delta$ (i.e. $n=0$), this contribution to the anomalous dimension is given by the inversion integral \cite{Sleight:2018epi}
\begin{equation}
    \frac12\gamma^{\text{n.-a.}}_{0,\ell|\tau,\ell}\,a^{(0)}_{0,\ell}=\int^{+i\infty}_{-i\infty}\frac{ds}{4\pi i}\,\Gamma\left(-\frac{s}{2}\right)^2\Gamma\left(\frac{s+2\Delta}{2}\right)^2\,M(s,2\Delta)\,Q_{\ell}^{(2\Delta,2\Delta,0,0)}(s)\,.
\end{equation}
The above integral can be evaluated by focusing on the term proportional to $s^\ell$, since all lower powers of $s$ are orthogonal to $Q_{\ell}^{(2\Delta,2\Delta,0,0)}(s)$. This term reads 
\begin{align}
    M(s,t)&=\sum^\infty_{m=0}M_m(s,t)\\
    M_m(s,t)&=s^\ell\left[\tfrac{(-1)^{\ell+m+1} 2^{\ell+\tau -3} \Gamma \left(\ell+\frac{\tau +1}{2}\right) (-2 \Delta +2 m+\tau ) \Gamma (d-\ell-\tau -1) \Gamma \left(\frac{d-2 (\ell+m+\tau )}{2}\right)}{\sqrt{\pi } \Gamma (m+1) \Gamma \left(\frac{\tau }{2}\right)^2 \Gamma (d-\tau -1) \Gamma \left(\ell+\frac{\tau }{2}\right) \Gamma \left(\frac{d}{2}-\ell-\tau \right) \Gamma \left(-m+\Delta -\frac{\tau }{2}+1\right)^2}\right.\\& \hspace*{4cm}\left.\times\,\frac{(\ell+\tau -1)_\ell (-d+\tau +2)_\ell}{2^\ell \left(\frac{\tau }{2}\right)_\ell^2}+O(t,s)\right]\,, \nonumber
\end{align}
which is independent of $t$. The sum over $m$ is given by a Gauss hypergeometric function evaluated at $z=1$:
\begin{equation}
    \, _2F_1\left(\begin{matrix}\tfrac{\tau -2 \Delta}{2},\tfrac{\tau -2 \Delta}{2}\\-\tfrac{d}{2}+\ell+\tau +1\end{matrix};1\right)=\frac{\Gamma \left(-\frac{d}{2}+\ell+2 \Delta +1\right) \Gamma \left(-\frac{d}{2}+\ell+\tau +1\right)}{\Gamma \left(-\frac{d}{2}+\ell+\Delta +\frac{\tau }{2}+1\right)^2}\,.
\end{equation}
The integral in $s$ therefore boils down to:
\begin{equation}
    \int^{+i\infty}_{-i\infty}\frac{ds}{4\pi i}\,\Gamma\left(-\frac{s}{2}\right)^2\Gamma\left(\frac{s+2\Delta}{2}\right)^2\,s^\ell\,Q_{\ell}^{(2\Delta,2\Delta,0,0)}(s)\,=\frac{(-4)^\ell \ell! \Gamma (\ell+\Delta )^4}{\Gamma (2 (\ell+\Delta )) (\ell+2 \Delta -1)_\ell},
\end{equation}
which, combined with the mean field theory OPE coefficient \eqref{MFTOPE}, gives the following expression for the non-analytic in spin contribution to the anomalous dimension $\gamma_{0,\ell|\tau,\ell}$,
\begin{equation}
    \gamma^{\text{n.-a.}}_{0,\ell|\tau,\ell}=\gamma^{\text{n.-a.}}_{0,0|\tau,0}\,\frac{2^{-\ell} \ell! (\tau+2\Delta-d ) (\Delta )_\ell \left(\frac{\tau +1}{2}\right)_\ell (\ell+\tau -1)_\ell \left(\frac{d}{2}-\ell-2 \Delta +1\right)_\ell \left(\frac{d-2 (\ell+\tau )}{2}\right)_\ell}{(2 (\Delta +\ell)+\tau -d)\left(\Delta +\frac{1}{2}\right)_\ell \left(\frac{\tau }{2}\right)_\ell^3 \left(\frac{-d+2 \Delta +\tau }{2}\right)_\ell^2}\,,
\end{equation}
where
\begin{equation}
    \gamma^{\text{n.-a.}}_{0,0|\tau,0}=\frac{2^{-2 \Delta +\tau +1} \Gamma (\Delta )^3 \Gamma \left(\frac{\tau +1}{2}\right) \Gamma \left(2 \Delta -\frac{d}{2}\right) \Gamma \left(-\frac{d}{2}+\tau +1\right)}{\Gamma \left(\Delta +\frac{1}{2}\right) \Gamma \left(\frac{\tau }{2}\right)^3 \Gamma \left(\Delta -\frac{\tau }{2}\right) \Gamma \left(\Delta -\frac{\tau }{2}+1\right) \Gamma \left(\Delta -\frac{d-\tau }{2}\right) \Gamma \left(\Delta -\frac{d-\tau}{2}+1\right)}\,.
\end{equation}
Using the projector \eqref{twistproj} one can systematically extract the anomalous dimensions of subleading $n>0$ double-twist operators in a similar way as detailed in \cite{Sleight:2018epi}, obtaining the following general formula:
\begin{multline}
     \gamma^{\text{n.-a.}}_{n,\ell|\tau,\ell}=\gamma^{\text{n.-a.}}_{0,0|\tau,0}\,\frac{(\tau -2 \Delta ) (d-2 \Delta -\tau )}{(2 \Delta +2 n-\tau ) (-d+2 \Delta +2 \ell+2 n+\tau )}\\\times\,\tfrac{2^{-\ell} \ell! \left(\frac{\tau +1}{2}\right)_\ell (\ell+\tau -1)_\ell \left(\frac{d}{2}\right)_{\ell+n} \left(\frac{d-2 (\tau +1)}{2}-\ell+1\right)_\ell \left(\frac{d-2 \Delta -1}{2}-n+1\right)_n (\Delta )_{\ell+n} \left(\frac{d-4 \Delta}{2}-\ell-n+1\right)_{\ell+n}}{n! \left(\frac{d}{2}\right)_\ell \left(\frac{\tau }{2}\right)_\ell^3 \left(\frac{d-2 \Delta -2}{2}-n+1\right)_n (d-n-2 \Delta )_n \left(\frac{2 \Delta +1}{2}\right)_{\ell+n} \left(\frac{2 \Delta +\tau-d}{2}\right)_\ell^2}\,.
\end{multline}
We remind the reader that the above results for the non-analytic in spin contributions to anomalous dimensions of spin-$\ell$ double-twist operators induced by a spin-$\ell$ exchange in the direct channel are universal for a given exchange amplitude, as they cannot be affected by the degree $s^{\ell-1}$ polynomial contact terms.

\end{appendix}

\bibliography{refs}
\bibliographystyle{JHEP}

\end{document}